\documentclass[pre, preprint, showpacs, superscriptaddress]{revtex4-1}
\usepackage{amsmath, graphicx,amssymb,color, bm}
 \setcitestyle{round}
\begin{document}
\title{Registered and antiregistered phase separation of mixed amphiphilic bilayers (condensed title:\ `Bilayer registration/antiregistration')}

\date{\today}

\author{J.~J.~Williamson}
\email{johnjosephwilliamson@gmail.com}
%\altaffiliation[Current address:\ ]{Department of Physics, Institute for Soft Matter Synthesis and Metrology, Georgetown University, 37th and O Streets, N.W., Washington, D.C. 20057, USA}
\affiliation{Department of Physics, Institute for Soft Matter Synthesis and Metrology, Georgetown University, 37th and O Streets, N.W., Washington, D.C. 20057, USA}
%\affiliation{Soft Matter Physics Group, School of Physics and Astronomy, University of Leeds, Leeds LS2 9JT, UK}
\author{P.~D.~Olmsted}
\email{pdo7@georgetown.edu}
%\altaffiliation[Permanent address:\ ]{Department of Physics, Institute for Soft Matter Synthesis and Metrology, Georgetown University, 37th and O Streets, N.W., Washington, D.C. 20057, USA}
\affiliation{Department of Physics, Institute for Soft Matter Synthesis and Metrology, Georgetown University, 37th and O Streets, N.W., Washington, D.C. 20057, USA}
%\affiliation{Soft Matter Physics Group, School of Physics and Astronomy, University of Leeds, Leeds LS2 9JT, UK}

\bibliographystyle{biophysj}
\renewcommand{\bibnumfmt}[1]{#1.}

%\pacs{87.16.dt, 87.16.dj, 87.14.Cc, 82.70.Uv}

\begin{abstract} 
We derive a mean-field free energy for the phase behaviour of coupled bilayer leaflets, which is implicated in cellular processes and important to the design of artificial membranes. Our model accounts for amphiphile-level structural features, particularly hydrophobic mismatch, which promotes antiregistration (AR), in competition with the `direct' trans-midplane coupling usually studied, promoting registration (R). We show that the phase diagram of coupled leaflets allows multiple \textit{metastable} coexistences, then illustrate the kinetic implications with a detailed study of a bilayer of equimolar overall composition. For approximate parameters estimated to apply to phospholipids, equilibrium coexistence is typically registered, but metastable antiregistered phases can be kinetically favoured by hydrophobic mismatch. Thus a bilayer in the spinodal region can require nucleation to equilibrate, in a novel manifestation of Ostwald's `rule of stages'. Our results provide a framework for understanding disparate existing observations, elucidating a subtle competition of couplings, and a key role for phase transition kinetics in bilayer phase behaviour.

\bigskip

\end{abstract}

\maketitle

\section{Introduction}

Phase separation in amphiphilic bilayers is of great interest due to cellular roles of lipid `rafts' \cite{Lingwood2010, Kusumi2004}, and as a means of designing function into artificial membranes. A full understanding of their rich phase behaviour requires consideration of the separate, yet coupled, leaflets of the bilayer \cite{May2009,Putzel2011,Funkhouser2013, Allender2006, Putzel2008, Wagner2007, Hirose2009}. Such inter-leaflet coupling is especially important in, e.g., mechanisms of protein localisation via lipid demixing \cite{Kusumi2004}. 

\begin{figure}[floatfix]
\includegraphics[width=8.25cm]{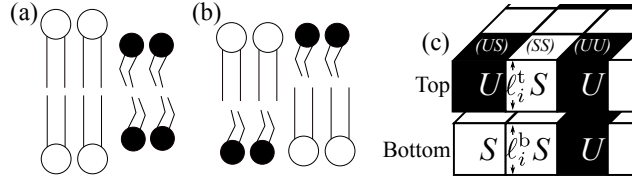}
\caption{\label{cartoon}
(a)~Registration (R), satisfying the direct coupling but with thickness mismatch penalised by the indirect coupling. (b)~Antiregistration (AR), without thickness variation but a mismatch penalty from the direct coupling. (c)~Lattice model for coupled bilayer leaflets. 
}
\end{figure}

Experiment and simulation yield disparate results. Observations of registered domains \cite{Korlach1999, *Dietrich2001, Collins2008} (Fig.~\ref{cartoon}a) imply a mismatch free energy per area favouring registration (R), which we call `direct' coupling \footnote{The mismatch free energy per area can be defined as $\gamma \equiv (G^\textrm{antireg}(A) - G^\textrm{reg}(A))/A$, where $G^\textrm{(anti)reg}(A)$ is the free energy of a large (anti)registered domain of area $A$.}. However, registered domains of different phases typically differ in hydrophobic thickness, arising from different preferred lengths of the mixed species due to differences in the molecular length and degree of ordering of their tails. In model phospholipid bilayers, typical measured thickness differences are between $\sim 0.2\!-\!1.6\,\textrm{nm}$ for liquid-ordered vs.\ liquid-disordered ($L_o$-$L_d$) lipid phases \cite{Garcia2007}, and slightly more for liquid-gel coexistence \cite{Lin2006}. Such `hydrophobic mismatch' can be alleviated by \textit{antiregistration} (AR, Fig.~\ref{cartoon}b);\ thus an `indirect' coupling favouring AR competes with the direct coupling. Antiregistration was inferred experimentally on the single-amphiphile level \cite{Zhang2004, *Zhang2007}, while AR \textit{domains} have appeared in $L_o$-$L_d$ \cite{Perlmutter2011} and liquid-gel \cite{Stevens2005, Bennun2007} simulations, and AFM on solid-supported bilayers has shown R gel domains decaying into AR \cite{Lin2006}.

Despite its wide practical importance, this complex behaviour lacks a full theoretical picture. Existing theories \cite{Putzel2008, May2009, Wagner2007} treat the bilayer as two phenomenologically coupled phase-separating leaflets, with an order parameter to describe the demixing transition. The phenomenological free energies and parameters in these models do not relate directly to any molecular or structural features of bilayers. Hydrophobic mismatch is often not explicitly included in coarse-grained modelling \cite{Sornbundit2014, May2009, Wagner2007, Putzel2008, Funkhouser2013, Hirose2009}, so that the competition of direct and indirect inter-leaflet couplings described above cannot be captured \footnote{Hydrophobic mismatch appears naturally in more complex molecular models \cite{Longo2009}. It was included in coarse-grained modelling in \cite{Wallace2006}, but without resolving the individual leaflets}. 

We approach the problem by deriving the bilayer's local free energy density from a lattice model of the coupled leaflets, in which simplified molecular interactions and bilayer structural features, including hydrophobic mismatch, appear explicitly. We show how competing interactions (favouring Fig.~\ref{cartoon}a, \ref{cartoon}b) lead to phase diagrams with multiple, competing coexistences. This implies competing modes of phase separation, and helps reconcile observations of registration and antiregistration in the literature \cite{Lin2006, Perlmutter2011, Stevens2005,Korlach1999, *Dietrich2001, Collins2008}.

As a test case, we study how antiregistration competes against registration for a bilayer containing an overall equimolar mixture of species in both leaflets. Antiregistration can become equilibrium, which arises from treating hydrophobic mismatch among individual molecules in the bulk (not only at domain boundaries), although most expected parameters yield equilibrium registration. However, metastable antiregistered states are still \textit{kinetically} favoured by hydrophobic mismatch. Hence, a bilayer in the conventional `spinodal' region can, paradoxically, require nucleation in order to reach equilibrium. Thus, metastable phases can interfere with bilayer domain registration and even prevent equilibration. 

\section{Theory}

\subsection{Lattice bilayer model}

To obtain the local free energy, we begin with a lattice model (Fig.~\ref{cartoon}c) for the two leaflets and their competing direct and indirect couplings. A local patch of the bilayer is modelled as an $L^2 = N$ square lattice of sites, where each site contains a `top' (t) and `bottom' (b) leaflet amphiphile. The lattice spacing is the lateral distance between amphiphiles, $a \sim 0.8\,\textrm{nm}$ for phospholipids. Each amphiphile has a hydrophobic length $\ell^\textrm{t(b)}_i$. We define the bilayer thickness $d_i \equiv \ell^\textrm{t}_i + \ell^\textrm{b}_i$, and the difference $\Delta_i \equiv \ell^\textrm{t}_i - \ell^\textrm{b}_i$. Extension of the tails also entails greater tail ordering, which we implicitly map onto the length variables $\ell^\textrm{t(b)}_i$ \cite{Komura2004}. We define $\hat{\phi}^\textrm{t(b)}_i = 1\textrm{ or } 0$ if the top (bottom) of site $i$ contains an `$S$' or `$U$' species amphiphile. These labels are chosen to suggest saturated and unsaturated lipids, where $S$ prefers a longer, more ordered tail structure;\ however they can represent any two species, or two of the lipid phases ($L_o$, $L_d$ or gel) available to a ternary ($S + U + \textrm{cholesterol}$) bilayer \cite{Sornbundit2014, Bagatolli2009, Honerkamp2008} \footnote{We use `phase' to refer to a bilayer phase, i.e.\ a given combination of the order parameters in each leaflet \cite{Putzel2008};\ `lipid phase' refers to particular ordering types ($L_o$, $L_d$, gel, etc.) which, in our model, are abstracted onto the binary $S$ and $U$ species \cite{Sornbundit2014, Bagatolli2009, Honerkamp2008}.}. Each lattice site may be `pairwise' registered ($SS$ or $UU$), or antiregistered ($SU$ or $US$). The species-dependent ideal hydrophobic lengths are $\ell_{S0},\,\ell_{U0}$. We define $\Delta_0 \equiv \ell_{S0} - \ell_{U0}$, $d_0 \equiv \ell_{S0} + \ell_{U0}$. We choose $\ell_{S0} > \ell_{U0}$, although this choice is arbitrary. 

The exact lattice Hamiltonian considered is
\begin{widetext}
\begin{align}\label{eqn:fun4}
H &= \sum_{<i,j>} ( V_{\hat{\phi}_i^\textrm{t} \hat{\phi}_j^\textrm{t}}  +  V_{\hat{\phi}_i^\textrm{b} \hat{\phi}_j^\textrm{b}}) 
+ \sum_{<i,j>} \tfrac{1}{2}\tilde{J} (d_i - d_j)^2 
+  \sum_{i} \tfrac{1}{2}B (\Delta_i)^2 
+  \sum_{i} \tfrac{1}{2}\kappa \left(( \ell^\textrm{t}_i - \ell_0^{\textrm{t}i})^2 + ( \ell^\textrm{b}_i - \ell_0^{\textrm{b}i})^2                      \right)~,
\end{align}
\end{widetext}
\noindent where $ \ell_0^{\textrm{t(b)}i} = \ell_{S0}$ for an $S$ amphiphile at the top (bottom) of site $i$, or $\ell_{U0}$ for $U$. 

The first two terms of $H$ are nearest-neighbour interactions. An Ising interaction $V_{uv}$ occurs among $S$ and $U$ amphiphiles within each leaflet separately, representing interactions independent of amphiphile length, such as those between headgroups. The hydrophobic penalty $\tilde{J}$ acts on the \textit{total} bilayer thickness, `indirectly' coupling the top and bottom amphiphiles of a given site via the surrounding thickness, and favours antiregistration to minimise thickness variation (Fig.~\ref{cartoon}b). The final two terms are on-site terms;\ $B$ is a direct coupling that favours registration (R), similarly to the conventional mismatch free energy density $\gamma$ \cite{Pantano2011, Risselada2008, *Polley2013, May2009, Putzel2011}, by penalising differences in length (thus tail ordering) between the top and bottom amphiphiles of a site. $\kappa$ penalises length stretching relative to the species' ideal lengths.

\subsection{Local free energy}\label{sec:localfreeenergy}

Our goal is the free energy per lattice site $f$, as a function of the coarse-grained local compositions $\phi^\textrm{t(b)} \equiv N_S^\textrm{t(b)} /N$, bilayer thickness $\bar{d} \equiv \sum d_i  /N$, and thickness difference $\overline{\Delta} \equiv \sum \Delta_i / N$. 
We calculate this within a mean-field approximation (Appendix~\ref{app:derivation}) in which the neighbour terms of Eq.~\ref{eqn:fun4} involving $V$ and $\tilde{J}$ are approximated by on-site terms.  
The coarse-grained local variables impose constraints 
\begin{subequations}
\begin{align}\label{eqn:sc1}
\sum_\alpha N_\alpha d_\alpha &= N \bar{d}~,\\
\label{eqn:sc2}
\sum_\alpha N_\alpha \Delta_\alpha &= N \overline{\Delta}~,\\
\label{eqn:sc3}
N_{SU} - N_{US} &= (\phi^\textrm{t} - \phi^\textrm{b}) N~,\\
\label{eqn:sc4}
N_{SS} + N_{SU} &= \phi^\textrm{t} N~,\\
\label{eqn:sc5}
N_{UU}+N_{US} &= (1-\phi^\textrm{t}) N~,
\end{align}
\end{subequations}
where $N_\alpha$ are the occupancies of the four possible site types $\alpha \in \{SS,UU,SU,US\}$ with thickness variables $d_\alpha,\,\Delta_\alpha$. After some work, we find (see Eq.~\ref{eqn:MFF2}) that the desired local free energy per-site $f (\phi^\textrm{t},\,\phi^\textrm{b},\,\bar{d},\,\overline{\Delta})$ is given by minimising 
\begin{align} \label{eqn:MFF}
f^{'} N = &\sum_\alpha (N_\alpha H_{\alpha} + k_\textrm{B}T N_\alpha \ln{N_\alpha}) 
- 2 V N (\phi^\textrm{t} - \phi^\textrm{b})^2 - 2 V  N(\phi^\textrm{t} + \phi^\textrm{b} -1)^2~,
\end{align}
\noindent subject to Eqs.~\ref{eqn:sc1}--\ref{eqn:sc5}, where $H_{\alpha}$ contain the thickness-dependent, mean-field interactions for each site type (cf.\ Eq.~\ref{eqn:Hlengthonly})
\begin{align}\label{eqn:MFHam}
H_{\alpha} = &\tfrac{1}{2}J (d_\alpha - \bar{d})^2 +\tfrac{1}{2} B (\Delta_\alpha)^2 
+\tfrac{1}{2}\kappa  \left(( \ell^\textrm{t}_\alpha - \ell_0^{\textrm{t}\alpha})^2 + ( \ell^\textrm{b}_\alpha - \ell_0^{\textrm{b}\alpha})^2                      \right)~,
\end{align}
\noindent in which $\ell_0^{\textrm{t}\alpha} = \ell_{A0}$, $\ell_0^{\textrm{b}\alpha} = \ell_{B0}$ for $\alpha = AB$, and $J \equiv 4 \tilde{J}$. $V  \equiv V_{10} - \frac{1}{2}(V_{11} +V_{00})$ sets the strength of the length-independent interaction. 

$f$ may be minimised over thickness variables $\bar{d}$ and $\overline{\Delta}$ to yield `annealed' values (equilibrated at given local composition) $\bar{d}^\textrm{[ann.]} = \Delta_0 (\phi^\textrm{t} + \phi^\textrm{b} -1) + d_0$ and $\overline{\Delta}^\textrm{[ann.]} = \kappa \Delta_0 (\phi^\textrm{t} - \phi^\textrm{b})/(2 B + \kappa)$. This gives the local free energy as a function of the local compositions of the top and bottom leaflets
\begin{align}
f^\textrm{[ann.]}(\phi^\textrm{t},\phi^\textrm{b}) \equiv  f (\phi^\textrm{t},\,\phi^\textrm{b},\,\bar{d}^\textrm{[ann.]},\,\overline{\Delta}^\textrm{[ann.]})~.
\end{align}
\noindent Explicit expressions are given in Eqs.~\ref{eqn:fullf} and \ref{eqn:fullfann}. Symmetry under exchange of labels $S$ and $U$ implies symmetry under $\phi^\textrm{t} \to 1-\phi^\textrm{t}, \phi^\textrm{b} \to 1-\phi^\textrm{b}$. Differing molecular area (for example) would break this symmetry but not affect the qualitative conclusions (in such a case `equimolar' should be read as `equal area fractions'). Identical material parameters within each leaflet imply symmetry under $\phi^\textrm{t} \to \phi^\textrm{b}, \phi^\textrm{b} \to \phi^\textrm{t}$;\ the qualitative effect of breaking this assumption is demonstrated in \cite{Putzel2008}.

We emphasise that $f$ describes the \textit{bulk} free energy of a local patch of bilayer. Hence, within $f$, neighbour interactions $V$ and $J$ penalise composition or thickness mismatch at the microscale, i.e.\ among individual amphiphiles in the local patch. $f$ does not include the contribution of boundaries between domains, which are irrelevant for large domains so do not affect phase equilibria. Their important effect on kinetics is studied in Section~\ref{sec:growthrates}. 

\section{Results}

We now study the implications of the free energy derived from our model. We will show how a particular local free energy landscape relates to the phase diagram of the system, and then study how model parameters affect the coexistences and instabilities governing an example bilayer in which each leaflet's overall composition is an equimolar mixture of $S$ and $U$.

\subsection{Parameters}

First, we discuss the estimated values/ranges and physical content of our model's interaction parameters $V$, $J$, $B$ and $\kappa$. Full details can be found in Appendix \ref{app:physical}.
The length-independent Ising interaction strength $V$ controls whether the leaflets would phase separate in the \textit{absence} of coupling ($J\!=\!B\!=\!0$, such that each leaflet acts as an independent Ising lattice). In the mean-field approximation, the Ising model requires $V  > V_{0} \equiv 0.5\,k_\textrm{B}T$ for phase separation, so we test values of $V$ above and below this threshold. 

The indirect coupling $\tilde{J}$ quantifies the penalty for mismatched total hydrophobic thickness. We take a fiducial value $\tilde{J} \approx 0.8\,k_\textrm{B}T\textrm{nm}^{-2}$, as estimated in \cite{Wallace2006} as a surface tension for hydrocarbon tails in contact with the watery headgroups of phospholipids. This gives $\tilde{J} \approx 0.5\,a^{-2}k_\textrm{B}T$, thus $J \approx 2\,a^{-2}k_\textrm{B}T$, and varying $J$ corresponds to varying the strength of hydrophobic mismatch/hydrophobicity. 

The direct coupling parameter $B$ plays a similar role to the inter-leaflet mismatch energy $\gamma$, for which widely varying estimates have been made \cite{May2009, Putzel2011, Risselada2008, *Polley2013}. The mechanism responsible is unclear. Proposals include tail interdigitation entropy \cite{May2009}, while \cite{Putzel2011} considers an interplay of entropic and enthalpic effects (such as tail orientation interactions and gauche bond energy) calculated from a molecular mean-field theory. Our specific choice of coupling $B$ to leaflet thickness, hence tail ordering, captures the idea that tail structural features underlie the direct coupling \cite{May2009, Putzel2011}, but does not qualitatively affect the results;\ it can simply be thought of as leading to an effective $\gamma$, which is plotted on Fig.~\ref{JBlines}. 

The stretching modulus $\kappa$ can be related to the \textit{area} stretching modulus $\kappa_{A}$ -- we use $\kappa = 3\,a^{-2}k_\textrm{B}T$, corresponding to $\kappa_{A} \approx 60\,k_\textrm{B}T\textrm{nm}^{-2}$, in the range for lipid bilayers at $300\,\textrm{K}$ \cite{Wallace2005, Needham1990, Rawicz2000}. Details of this mapping, as well as that from $B$ to $\gamma$, appear in Appendix~\ref{app:physical}.

Due to the simplicity of our model, precisely assigning the meaning and values of parameters is impossible;\ for example, amphiphiles could respond to length mismatch by exploring tilt and splay as well as the stretching modelled by our $\kappa$. Instead, our aim is to succinctly capture important structural features of the bilayer and study their effects over a range of reasonable estimates for the parameters involved. 

Given a quench into a phase separating region, we broadly expect increased direct coupling $B$ to penalise the existence of pairwise antiregistered sites $SU$ and $US$, while increased hydrophobic penalty $J$ will penalise the \textit{mixing} of sites with different ideal thickness ($SS$, $UU$ and $SU$/$US$). Varying stiffness $\kappa$ affects the characteristic energy scale of the inter-leaflet couplings;\ $\kappa \to 0$ would represent infinitely `floppy' amphiphiles which can adjust their length and structuring so as to experience no indirect or direct coupling energy.

\subsection{Phase diagram}

\begin{figure}[floatfix]
\includegraphics[width=8.25cm]{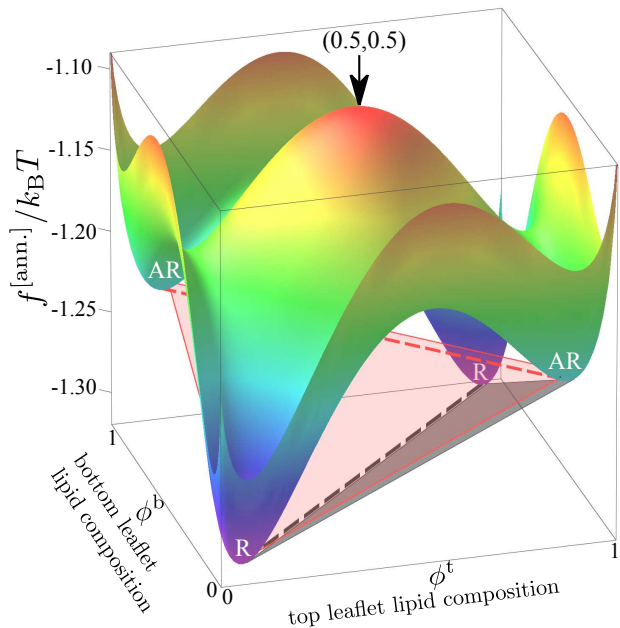}
\caption{\label{botleft}Local free energy landscape for the parameter point marked in Fig.~\ref{JBlines}c. The AR-AR and R-R central tie-lines (dashed) of Fig.~\ref{phasediag} are superimposed, along with two illustrative tangent planes corresponding to three-phase triangles (equilibrium R-R-AR black, metastable AR-AR-R red). 
}
\end{figure}

\begin{figure}[floatfix]
\includegraphics[width=8.25cm]{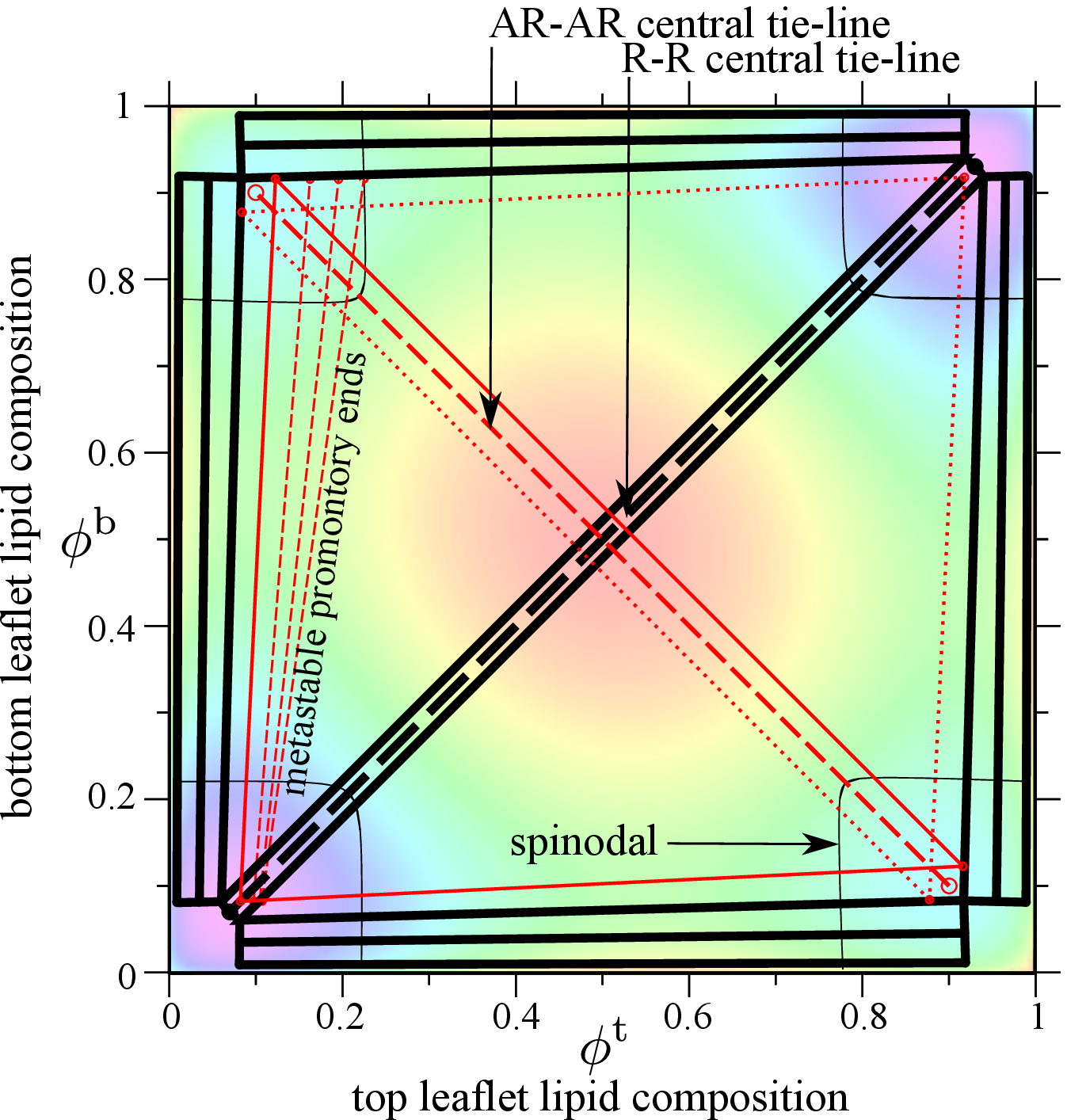}
\caption{\label{phasediag}Phase diagram calculated from the local free energy landscape in Fig.~\ref{botleft}. Thick black tie-lines or triangles show equilibrium two- or three-phase coexistence. Thin red lines or triangles mark metastable two- or three-phase coexistence. One of the overlapping AR-AR-R triangles is dotted for clarity. A two-phase promontory is shown, other metastable two-phase regions are omitted. Thin black lines mark the spinodals.
}
\end{figure}

Our $f^\textrm{[ann.]}(\phi^\textrm{t},\phi^\textrm{b})$ plays the same role as the local free energies postulated in \cite{Putzel2008, Wagner2007}, except that we have derived it explicitly from the lattice model. Similarly, it can be used to find a phase diagram \cite{Putzel2008, Wagner2007}, which we now perform for the particular local free energy landscape shown in Fig.~\ref{botleft}. Hereafter we assume no flip-flop or solvent exchange, so the overall composition (total proportion of $S$ and $U$) within each leaflet is conserved. The total free energy $F$ is found by integrating $f$ laterally over the entire bilayer \cite{Wagner2007}. The equilibrium state -- which coexisting phases are present at equilibrium -- is then determined by minimising $F$ subject to constraints specifying the overall leaflet compositions
\begin{align}\label{eqn:constraints}
\sum^\textrm{phases} \theta_n &= 1~,\notag\\
\sum^\textrm{phases} \theta_n \phi^\textrm{t}_n &= \Phi^\textrm{t}~,\notag\\
\sum^\textrm{phases} \theta_n \phi^\textrm{b}_n &= \Phi^\textrm{b}~,
\end{align}
\noindent where $\theta_n$ label the coexisting phases' area fractions. The overall (conserved) leaflet compositions in Eq.~\ref{eqn:constraints} are labelled $\Phi^\textrm{t,b}$, but by convention it is unnecessary to introduce a new symbol as the distinction between local and overall compositions is clear from context \cite{Wagner2007}. 

To stably (or metastably) coexist, phases must have i)~equal chemical potential $\mu^\textrm{t} \equiv \partial f / \partial \phi^\textrm{t}$ in the top leaflet, ii)~equal chemical potential $\mu^\textrm{b} \equiv \partial f / \partial \phi^\textrm{b}$ in the bottom leaflet and iii)~equal surface tension $f -\mu^\textrm{t}\phi^\textrm{t} - \mu^\textrm{b}\phi^\textrm{b}$ \cite{Putzel2008}. This is equivalent to drawing common tangent \textit{planes} touching the surface $f^\textrm{[ann.]}(\phi^\textrm{t},\phi^\textrm{b})$ at two or three points, which define two-phase tie-lines or triangles of three-phase coexistence. This concept is illustrated on Fig.~\ref{botleft}.

The phase diagram derived from Fig.~\ref{botleft} is shown in Fig.~\ref{phasediag}. Equilibrium coexistences are qualitatively identical to those in \cite{Putzel2008, Wagner2007}. Spinodal lines around each free energy minimum indicate the region of local stability \cite{Wagner2007}. The registered `central' tie-line runs along $\phi^\textrm{b}_\textrm{R}(\phi^\textrm{t}) = \phi^\textrm{t}$ through $(0.5,0.5)$, linking the registered minima of Fig.~\ref{botleft}. (Hereafter, we use $(0.5,0.5)$ as shorthand for $\phi^\textrm{b}\!=\!\phi^\textrm{t}\!=\!0.5$.) It sits within a region of two-phase R-R equilibrium. This is surrounded by triangles of three-phase R-R-AR equilibrium, where two registered phases coexist with one antiregistered phase. These connect to two-phase `arms' of R-AR coexistence.

We also show some \textit{metastable} coexistences, which satisfy the common tangent condition, but do not fully minimise $F$. Metastable AR-AR-R triangles overlap one another, and a central AR-AR tie-line runs along $\phi^\textrm{b}_\textrm{AR}(\phi^\textrm{t}) = 1-\phi^\textrm{t}$. Each pair of free energy minima is associated with a metastable `promontory' that encroaches into an equilibrium three-phase region. This is illustrated for one promontory on the figure;\ a similar idea applies to each pair of minima.
Tie-line endpoints must not be unstable, hence the spinodals determine where these promontories end. Carefully inspecting Fig.~\ref{phasediag} reveals the following equilibrium regions:\ two three-phase (R-R-AR);\ five two-phase (four R-AR plus one R-R). The metastable regions are:\ two three-phase (AR-AR-R);\ six two-phase (four R-AR, one AR-AR and one R-R). 
Metastable states, unlike equilibria, are not uniquely defined for each point on the phase diagram. For example, a state point near an AR minimum could lie within \textit{both} metastable three-phase triangles, and \textit{three} distinct two-phase promontories. 

Metastable coexistences are not restricted to the free energy derived here, but apply to any landscape containing both registered and antiregistered minima. 
Such free energies have been used to explain existing observations \cite{Collins2008, Putzel2008}, suggesting that bilayer free energies may generically permit the metastable coexistences identified here. The free energy and phase diagram are symmetric under inversion through $(0.5,0.5)$ for the reasons outlined in Section~\ref{sec:localfreeenergy}. (For a general free energy landscape without these symmetries, coexisting phases need not be minima so long as points of inflection exist. In our symmetric case, R-R-AR could occur without AR minima, but coexistence of two AR phases requires AR minima.) 
Breaking the `up-down' leaflet symmetry (e.g.\ one leaflet containing different species $S'$, $U'$) could be treated by modifying the Hamiltonian. The qualitative effects would resemble the case in \cite{Putzel2008} where a different intra-leaflet parameter is used for each leaflet.

\section{Kinetics for $\phi^\textrm{b}\!=\!\phi^\textrm{t}\!=\!0.5$}

The importance of metastable states in determining the kinetics of realised phase behaviour has long been known in the metallurgy and colloid literatures \cite{Ostwald, Poon1999, Poon2002}, but until now has not been discussed for bilayer leaflets. 
We now show these kinetic implications by studying an example bilayer whose leaflets each contain an equimolar mixture of $S$ and $U$ (marked $(0.5,0.5)$ in Fig.~\ref{botleft}). Immediately after a quench from high temperature its \textit{local} composition will everywhere be homogeneous at $(0.5,0.5)$. Varying model parameters (hence the free energy landscape) we consider:\ i)~What is its equilibrium state? ii)~Which instabilities is the initial homogeneous state subject to and, if more than one, which will dominate at the start of phase separation? 

For $(0.5,0.5)$ overall composition the equilibrium state must, if phase separated, be two-phase, since any three-phase tangent plane would pass through $(0.5,0.5)$ at a higher free energy than a plane linking the two absolute minima of $f$. 
For the initial homogeneous state, we compare the R phase separation mode (in which the bilayer splits in the direction of the R-R central tie-line) against the AR mode (splitting along the perpendicular AR-AR tie-line), ignoring the metastable three-phase triangles in which $(0.5,0.5)$ also falls. Restricting attention to the two perpendicular modes simplifies the kinetic analysis. Simulations to be presented in upcoming work suggest that, for $(0.5,0.5)$ overall composition, metastable three-phase separation is either kinetically disfavoured or occurs only in small transient fluctuations about an overall AR-AR state. 

\begin{figure}[floatfix]
\includegraphics[width=8.25cm]{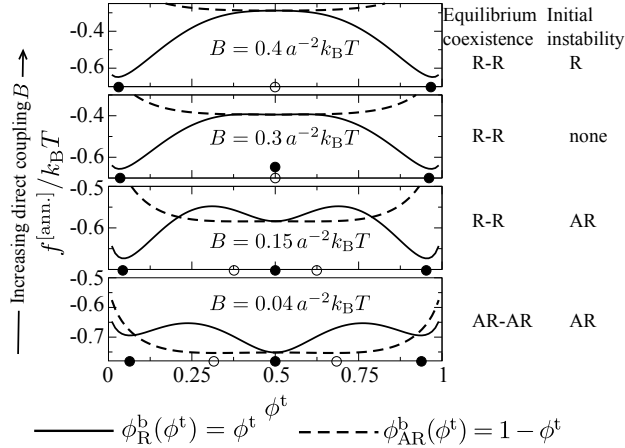}
\caption{\label{surfaces}R and AR slices through $f^\textrm{[ann.]}(\phi^\textrm{t},\,\phi^\textrm{b})$ for the sequence of parameter points marked in Fig.~\ref{JBlines}b. Filled (empty) circles mark minima in the R (AR) direction. We label, for a bilayer of overall composition $(0.5,0.5)$, whether equilibrium coexistence is R-R or AR-AR, and which modes (R or AR) the initially uniform homogeneous state is unstable to.} 
\end{figure}

\subsection{Registered and antiregistered modes}

Fig.~\ref{surfaces} shows slices through $f^\textrm{[ann.]}(\phi^\textrm{t},\,\phi^\textrm{b})$ in the R/AR modes, for different parameter points (marked in Fig.~\ref{JBlines}b) of varying direct coupling $B$. For the lowest $B$, the AR minima are lower than R, so AR-AR phase separation is the \textit{equilibrium} state of our $(0.5,0.5)$ bilayer. This arises because, in a registered domain of $SS$ (say), the entropic gain of inserting minority sites $SU$, $US$, or $UU$ is offset by a prohibitive hydrophobic cost from $J$. In contrast, an antiregistered $SU$ domain can gain entropy from minority $US$, which are of the same thickness so experience no hydrophobic penalty. Thus, if $J$ is large, the bulk free energy for AR-AR coexistence can be lower than R-R, despite a finite direct coupling $B$. 

Upon increasing $B$ (penalising AR), R-R phase separation becomes equilibrium, but the homogeneous state $(0.5,0.5)$ is a local minimum of $f$ along $\phi^\textrm{b}_\textrm{R}(\phi^\textrm{t}) = \phi^\textrm{t}$, so is metastable against the R mode. At $B=0.3\,a^{-2}k_\textrm{B}T$, the minima in the AR mode have disappeared, and the homogeneous state is not unstable to either mode. For the largest $B$ the homogeneous state becomes \textit{unstable} to R-R phase separation, the local minimum having disappeared. Note that, unlike any of the parameter points in Fig.~\ref{surfaces}, the free energy landscape in Fig.~\ref{botleft} is unstable at $(0.5,0.5)$ to \textit{both} R and AR modes of phase separation, being concave down along both $\phi^\textrm{b}_\textrm{R}(\phi^\textrm{t}) = \phi^\textrm{t}$ and $\phi^\textrm{b}_\textrm{AR}(\phi^\textrm{t}) =1 - \phi^\textrm{t}$ directions. 

\subsection{Instability criteria} \label{sec:criteria}

Instability to phase separation requires negative curvature of $f^\textrm{[ann.]}$. We define $f^\textrm{m[ann.]}(\phi^\textrm{t}) \equiv f^\textrm{[ann.]}(\phi^\textrm{t},\phi^\textrm{b}_\textrm{m}(\phi^\textrm{t}))$ where $\textrm{m} = \textrm{R, AR}$ labels the phase separation mode. Hence, instability of the equimolar `homogeneous state' to mode $\textrm{m}$ requires $ \left.\tfrac{d ^2}{{d \phi^\textrm{t}}^2} f^\textrm{m[ann.]} \right\rvert_{\phi^\textrm{t} = 0.5} < 0$. We find
\begin{subequations}\label{eqn:criteria}
\begin{align}
&\left.\tfrac{d ^2}{{d \phi^\textrm{t}}^2} f^\textrm{R[ann.]}\right\rvert_{0.5} = \frac{4}{\beta} ( 1+e^{-\beta \sigma}) - 16\!\left(\!V + \frac{\Delta_0^2 J \kappa}{4 (2J + \kappa)}\right)\label{eqn:instabR}~,\\
&\left.\tfrac{d ^2}{{d \phi^\textrm{t}}^2} f^\textrm{AR[ann.]}\right\rvert_{0.5} = \frac{4}{\beta}( 1+e^{\beta \sigma}) - 16 V\label{eqn:instabAR}~,
\end{align}
\end{subequations}
\noindent with $\beta^{-1} \equiv k_\textrm{B}T$, where
\begin{subequations} \label{eqn:X}
\begin{align}
\sigma &\equiv \tfrac{1}{2}(H_{SU} + H_{US} - H_{SS} -H_{UU})  \\
&= - \frac{\Delta_0^2 \kappa^2 (J-B)}{2(2 J + \kappa) (2 B + \kappa)}~,
\end{align}
\end{subequations}
\noindent is the (mean-field) energy per site for converting two sites from R to AR ($SS + UU \to SU+US$). The curvatures contain a positive entropy-like part inhibiting instability and a negative enthalpic part promoting it. The more negative is $\tfrac{d ^2}{{d \phi^\textrm{t}}^2} f^\textrm{m[ann.]}$, the stronger is the bulk driving force for instability to mode $\textrm{m}$.

$\sigma$ controls the excess proportion $x_\textrm{reg}^\textrm{mixed}$ of pairwise R amphiphile pairs present in the homogeneous state \footnote{While laterally mixed, the bilayer can contain more or fewer \textit{pairwise} R vs.\ AR sites. At $\phi^\textrm{b}\!=\!\phi^\textrm{t}\!=\!0.5$ in particular, anything from full pairwise R to AR is possible.}. Evaluating the $N_\alpha$ at $(0.5,0.5)$ gives
\begin{align} \label{eqn:delreg}
x_\textrm{reg}^\textrm{mixed} &\equiv \frac{N_{SS} + N_{UU} - N_{SU} -N_{US}}{N}
= \tanh{\frac{\beta \sigma}{2}}~.
\end{align}
\noindent For $J>B$, as expected in most cases, $\sigma < 0$ which implies $x_\textrm{reg}^\textrm{mixed}  < 0$. This implies most pairs in the homogeneous state are AR, i.e.\ amphiphiles predominantly align with the opposite species (as measured in \cite{Zhang2004,*Zhang2007}). The Boltzmann factors $e^{\pm\beta\sigma}$ in Eqs.~\ref{eqn:instabR} and \ref{eqn:instabAR} control the loss of configurational entropy, relative to the homogeneous state, for creating excess pairwise R (AR) sites required by the R (AR) phase separation mode.
Increasing $J$ (decreasing $\sigma$) promotes pairwise AR (i.e.\ $SU$ and $US$), by penalising the mixing of pairwise R ($SS$ and $UU$) sites in the homogeneous state. To access the R phase separation mode, the bilayer must thus overcome a free energy barrier to create the required pairwise R sites. This can lead to a local minimum at $(0.5,0.5)$ in the registered slice through $f$ (see Fig.~\ref{surfaces}), implying the homogeneous state is metastable against the R mode \footnote{The homogeneous state may become metastable against the AR mode if $B \gg J$, but for phospholipids the literature suggests $J \gtrsim B$ or $J \gg B$ (see Appendix~\ref{app:physical}). Further, the `complementary matching' measured in \cite{Zhang2004,*Zhang2007} requires, within our model, $J > B$ (via Eq.~\ref{eqn:delreg}).}. 
Hence hydrophobicity $J$, unlike the Ising interaction $V$, does not trivially increase instability to the R mode.
This complex interplay with bilayer microstructure cannot be captured in theories which a priori assume purely inter- and intra-leaflet couplings \cite{Putzel2008, May2009}.  

\subsection{Growth rates of competing modes}\label{sec:growthrates}

If the initial homogeneous state is unstable to \textit{both} R and AR modes, initial phase separation will be determined by the competition between them. The bulk free energy $f$ drives separation into domains, while gradient terms arising from the nearest-neighbour interactions $V$ and $J$ penalise the resulting inhomogeneities in composition and thickness. Although they do not affect the phase diagram, these gradient terms do affect the growth rates of the competing modes. We employ linear stability analysis of a Ginzburg-Landau (G-L) free energy \cite{Chaikin} $F_\textrm{G-L} = \int d^2 r \left((f/a^2) + f_\textrm{grad} \right)$, with gradient terms given by
\begin{align}
f_\textrm{grad} = \tfrac{1}{2}\tilde{J}(\nabla \bar{d})^2 + {V}(\nabla \phi^\textrm{t})^2 + {V} (\nabla \phi^\textrm{b})^2~.
\end{align}
\noindent We obtain wavenumber dependent growth rates $\omega^\textrm{m}(q)$ whose maxima over $q$ yield $\omega^\textrm{m}_\textrm{max}$. The difference $\Delta \omega \equiv \omega^\textrm{R}_\textrm{max} - \omega^\textrm{AR}_\textrm{max}$ determines which mode is faster and dominates the initial phase separation after a quench. Hydrophobic mismatch penalises the R mode's thickness gradients without necessarily providing a compensating boost to instability (Section~\ref{sec:criteria}), so can render the AR mode fastest. The detailed calculations are given in Appendix~\ref{app:ginzburg}. 

\subsection{Stability diagrams}\label{sec:stabdiagrams} 

\begin{figure*}[floatfix]
\includegraphics[width=16.5cm]{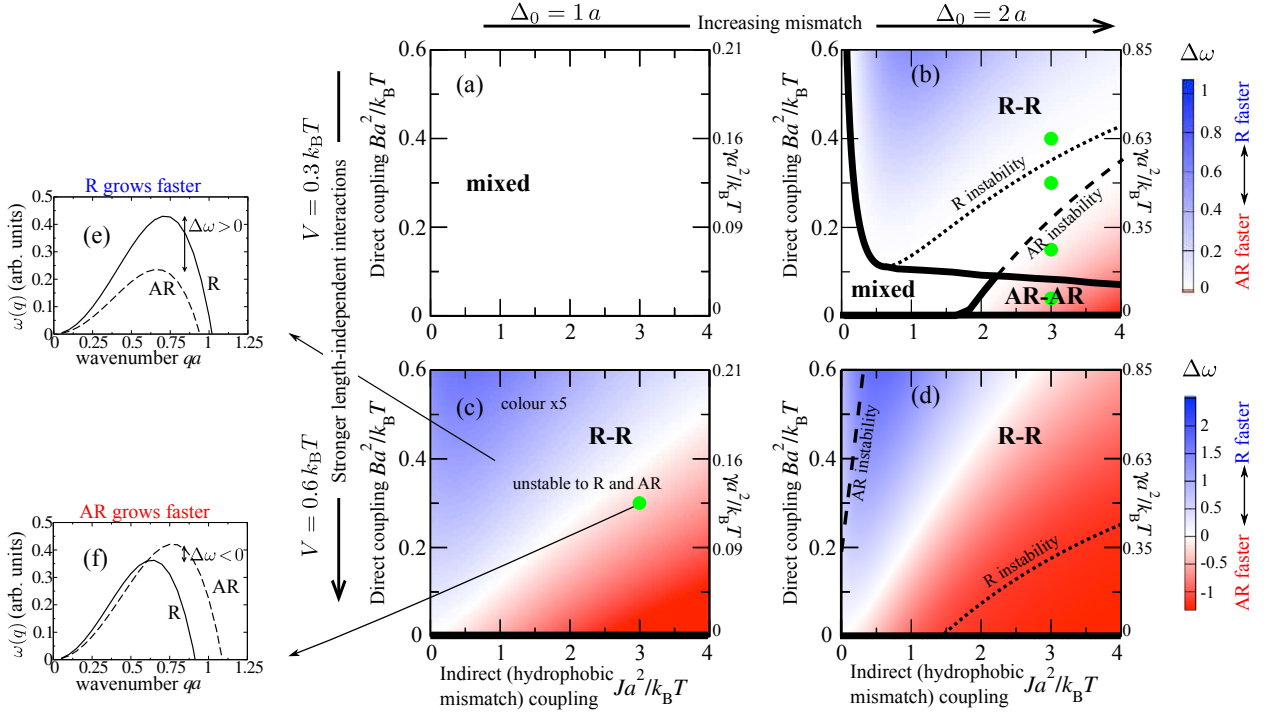}
\caption{\label{JBlines}(a)--(d)~Stability diagrams for $(0.5,0.5)$ overall composition, for varying indirect and direct couplings $J,\,B$, and values of the length-independent interaction strength $V$ and mismatch $\Delta_0$. Thick lines and bold labels indicate where the equilibrium state is R-R, AR-AR or mixed. Thin lines denote where the bilayer is unstable to R/AR modes, labels on the side of the lines to which they refer. \textbf{Secondary axis:}~approximate values of the inter-leaflet mismatch free energy per area $\gamma$ (Eq.~\ref{eqn:mismatchenergy}). \textbf{Colours:}~growth rates from linear stability analysis of the initial homogeneous state, $\Delta \omega > 0$ (R mode faster, blue), $\Delta \omega < 0$ (AR faster, red). Different colour scales are used for visibility, since comparison of growth rates between different panes of the figure has little meaning. In (c) the colour range is reduced $5$ times from that indicated on (d). \textbf{Green dots:}~in (b) correspond to Fig.~\ref{surfaces}, in (c) corresponds to Fig.~\ref{botleft}.  (e)--(f)~Illustrative $q$-dependent growth rates. 
}
\end{figure*}

Fig.~\ref{JBlines} summarises the equilibria and kinetics of a $(0.5,0.5)$ bilayer, showing whether the equilibrium state is R-R, AR-AR, or mixed (no phase separation at all). We have also shown where the bilayer's initial homogeneous state is unstable to R and AR modes, and their relative growth rates.

\noindent \textit{{Weak mismatch }}($\Delta_0 = 1\,a$):\ For $V = 0.3\,k_\textrm{B}T$ and weak thickness mismatch, no phase separation takes place since $V < V_0$ (Fig.~\ref{JBlines}a), while $V = 0.6\,k_\textrm{B}T$ (Fig.~\ref{JBlines}c) induces phase separation as for the mean-field Ising model. The equilibrium coexistence is R-R, but the bilayer is unstable to both R and AR modes. For strong enough hydrophobic mismatch, the AR mode is faster (red).
Hence, for the parameter point marked in Fig.~\ref{JBlines}c the bilayer will initially undergo spinodal decomposition in the AR mode, accessing metastable AR-AR coexistence, and subsequently requiring \textit{nucleation} to reach equilibrium R-R coexistence.

\noindent \textit{{Strong mismatch }}($\Delta_0 = 2\,a$):\ Increasing $\Delta_0$ strengthens both the indirect and direct couplings (physically, this could arise from increasing the length mismatch and difference in unsaturation of the species' tails). In contrast to the weak mismatch case, for $V = 0.3\,k_\textrm{B}T$ (Fig.~\ref{JBlines}b) the inter-leaflet couplings induce phase separation although (since $V <V_{0}$) \textit{neither} leaflet would separate without inter-leaflet coupling \cite{May2009, Wagner2007, Putzel2008}. A large hydrophobic penalty $J$ promotes pairwise AR. Due to the doubled effective Ising interaction between $SU$ and $US$ pairs ($2 V >  V_{0}$), AR minima appear in the free energy landscape and AR-AR phase separation is possible. There is a region where AR-AR coexistence is the \textit{equilibrium} state \footnote{A small region of AR-AR equilibrium also exists on Fig.~\ref{JBlines}c, d, for $Ba^2/k_\textrm{B}T \lesssim 0.005$}. Increasing the direct coupling $B$ favours R-R phase separation, which is enhanced by hydrophobic thickness mismatch between $SS$ and $UU$ sites -- yet large $J$ renders the homogeneous state metastable, not unstable, against the R mode (within the \textbf{R-R} region but outside the `R instability' line). For $V = 0.6\,k_\textrm{B}T$ (Fig.~\ref{JBlines}d), phase separation always takes place since $V >  V_{0}$. Compared to $V = 0.3\,k_\textrm{B}T$, the `R instability' and `AR instability' lines move past each other;\ increased $V$ promotes instability (Eq.~\ref{eqn:criteria}), so larger $B$/smaller $J$ is required to inhibit AR instability, and larger $J$/smaller $B$ is required to inhibit R instability.

\section{Discussion}

We have modelled the coupled leaflets of a bilayer in which hydrophobic mismatch causes an indirect coupling $J$, promoting antiregistration (AR). This competes with a direct coupling $B$, arising from tail structure mismatch, which promotes trans-midplane registration (R) of like species. Both inter-leaflet couplings interplay with the stiffness $\kappa$:\ small $\kappa$ allows amphiphiles to adapt to the couplings, decreasing the energy scale and `washing-out' inter-leaflet coupling effects, while large $\kappa$ (as in a gel, for instance) would strengthen them. 

The free energy landscapes derived from our model permit multiple metastable coexistences in the phase diagram. Such coexistences are possible for \textit{any} free energy landscape with registered and antiregistered minima \cite{Collins2008, Putzel2008}, but their consequences for bilayers have not previously been investigated. Moreover, by explicitly incorporating structural features, our theory demonstrates how hydrophobic mismatch kinetically favours metastable antiregistered phase coexistence (or can even lead to \textit{equilibrium} antiregistration), thus providing a novel link between bilayer microstructure and phase transition kinetics.

We demonstrated the kinetic effects of metastability in a $(0.5,0.5)$ bilayer (each leaflet containing an overall equimolar mixture), by studying competing R and AR phase separation modes corresponding to perpendicular tie-lines passing through $(0.5,0.5)$ (see Fig.~\ref{phasediag}). 
For plausible phospholipid parameters ($J \sim 2\, a^{-2}k_\textrm{B}T$) with a significant lipid length mismatch $\Delta_0 \sim 0.8\,\textrm{nm}$ ($ \sim 1\,a$), Fig.~\ref{JBlines}c may apply. Taking $\gamma \approx 0.15\,k_\textrm{B}T\textrm{nm}^{-2} \approx 0.1\,a^{-2}k_\textrm{B}T$ \cite{May2009, Risselada2008, *Polley2013} (so $B \approx 0.23\,a^{-2}k_\textrm{B}T$) on that figure implies comparable R/AR growth rates. However, \cite{Putzel2011} argues for $\gamma \sim 0.01\,a^{-2}k_\textrm{B}T$ ($B \approx 0.02\,a^{-2}k_\textrm{B}T$);\ then, the bilayer would first access AR-AR coexistence and require nucleation to reach equilibrium R-R coexistence (a manifestation of Ostwald's heuristic `rule of stages' \cite{Ostwald}). 

A general experimental signature of such kinetics would be the total amounts of registered and antiregistered phases changing through time, as the bilayer converts antiregistration to registration or vice versa \cite{Lin2006}. This could be discerned via AFM, with growing registered $SS$ and $UU$ nuclei exhibiting different thickness to one another as well as to the surrounding antiregistered background. This signature applies also to overall compositions away from $(0.5,0.5)$, where metastable states can compete with equilibrium three- or two-phase coexistence. 
AR-AR coexistence may not be detected in standard height-mode AFM or fluorescence because $SU$ and $US$ domains would be of similar height and fluorescence. Hence, \textit{three}-phase coexistence involving two AR phases can masquerade as two-phase coexistence. 

Observations of $L_o$-$L_d$ phase coexistence \cite{Korlach1999, *Dietrich2001, Collins2008} suggest that the equilibrium state typically comprises registered domains, in agreement with our results. In contrast, AFM experiments \cite{Lin2006} have shown registered gel domains converting to antiregistration. This could indicate decay of subcritical R nuclei into a metastable AR state, although the interpretation of experiments with a solid support is complicated by substrate effects \cite{Garg2007} which could break the bilayer symmetry. Domain antiregistration observed upon increasing hydrophobic mismatch in simulation \cite{Perlmutter2011, Stevens2005} can be understood as a kinetically favoured metastable state, and thus reconciled with registration as the equilibrium state \cite{Korlach1999, *Dietrich2001, Collins2008}. The intriguing `complementary matching' (i.e.\ \textit{pairwise} antiregistration) measured in \cite{Zhang2004, *Zhang2007} can be related to Eq.~\ref{eqn:delreg}, which implies (for $J > B$) predominant pairwise AR in a laterally homogeneous bilayer.   

To study these effects in experiment and molecular simulation, ideal systems would have strong hydrophobic mismatch due to different tail lengths. Differing headgroups may ensure that the length-independent interaction $V$ is sufficient for AR free energy minima to exist (so $SU$ and $US$ form distinct domains). Although metastability is possible over wide regions of the phase diagram, a mixture of near-equal area fractions (near $(0.5,0.5)$) could minimise bias towards registered phases and allow `pure' AR-AR coexistence, cf. \cite{Perlmutter2011}. Amphiphiles with relatively stiff tails (corresponding to a large area compressibility modulus) would maximise the energy scale of inter-leaflet couplings. The behaviour within the individual leaflets should be carefully monitored \cite{Collins2008, Garg2007}, ideally in the early kinetics after a quench where one might witness the bilayer passing through metastable states. Note, however, that some existing simulation \cite{Perlmutter2011, Stevens2005} and experimental results imply long-lived antiregistration. These could provide useful starting points for investigations, e.g.\  molecular simulations aimed explicitly at determining whether antiregistered domains in \cite{Perlmutter2011} are stable and how they may transition to registration. 

Notwithstanding quantitative effects on the parameters, our theory applies equally to $L_o$-$L_d$ or liquid-gel systems;\ gels entail slower kinetics but still evolve through time \cite{Lin2006} in a manner governed by a free energy landscape. In our idealised treatment it is hard to precisely assign parameters or predict timescales. Rather, we have shown that bilayer structural features lead -- over a wide range of reasonable model parameters -- to uniquely rich free energies that can reconcile apparently contradictory registration/antiregistration observations already present in the literature. Future work will flesh out the kinetics beyond the linear regime studied here by direct simulation of the lattice model, and examine the nucleation energetics for reaching equilibrium.

We have implicitly considered a flat membrane. Membrane curvature can drive domain formation \cite{Shlomovitz2013} and has been proposed as a factor in domain inter-leaflet coupling \cite{Pantano2011}, while antiregistration may cause nonzero spontaneous curvature and lead to an undulating membrane \cite{Funkhouser2013}. In principle our theory could be supplemented with curvature terms \cite{Funkhouser2013}, though it is not obvious whether these could be derived from microscopic considerations, or would be phenomenological. We have focused on approximate phospholipid parameters, but the phenomenology also applies to, e.g., polymeric bilayers \cite{Lee2012,Schulz2011}, whose properties and hence predicted phase behaviour may be quite different. `Hybrid' lipids, where one tail varies in saturation/length relative to the other, may act as linactants \cite{Brewster2009, Schafer2010, Palmieri2014}. Further work is required to establish their effects on the physics studied here, but we speculate that such linactants could favour registration, by diminishing the energy cost for the thickness mismatch at registered domain boundaries.

\appendix

\section{\label{app:derivation}Derivation of mean-field free energy}
The underlying lattice model is a $L^2 = N$ square lattice of sites with top (t) and bottom (b) leaflet amphiphiles, whose Hamiltonian is given by Eq.~\ref{eqn:fun4}. The exact partition function is 
\begin{align}
Z = \sum^\textrm{constrained}_{\{\hat{\phi}_i^\textrm{t},\hat{\phi}_i^\textrm{b}\}} \int \mathcal{D}\Delta \mathcal{D} d \exp{(-\beta H)}~,
\end{align}
\noindent where the sum is constrained by the average leaflet compositions $\phi^\textrm{t(b)} \equiv N_S^\textrm{t(b)}/N$, and we have defined $\mathcal{D} \Delta \equiv \prod_i d \Delta_i$ and $\mathcal{D} d \equiv \prod_i d d_i$. Recall $\hat{\phi}_i^\textrm{t(b)} = 1$ or $0$ if the top (bottom) of site $i$ contains an $S$ or $U$ amphiphile.
The free energy is related to the partition function by
\begin{align} \label{eqn:bythis}
fN = -k_\textrm{B}T\ln{Z}~,
\end{align}
\noindent and our aim is to find a mean-field approximation to the free energy per site $f$ that depends only on local coarse-grained variables for the leaflet compositions $\phi^\textrm{t(b)}$, bilayer thickness $\bar{d} \equiv \sum d_i  /N$, and thickness difference $\overline{\Delta} \equiv \sum \Delta_i / N$.

\subsection{Mean-field (on-site) approximation}

We use a mean-field approximation, ignoring correlations between neighbouring sites. This requires approximating the neighbour interaction terms of $H$ (those involving $V$ and $\tilde{J}$) with on-site terms. 
For the $\tilde{J}$ term, we employ the local mean-field bilayer thickness $\bar{d}$ and write 

\begin{align}
&\sum_{<i,j>}\left(d_i - d_j\right)^2 
=\sum_{<i,j>}\left([d_i - \bar{d}] - [d_j -\bar{d}]\right)^2 \notag \\
&=\sum_{<i,j>}\left([d_i - \bar{d}]^2 + [d_j - \bar{d}]^2 - 2[d_i - \bar{d}][d_j - \bar{d}]\right) \notag \\
&\approx \sum_i 4 ( d_i - \bar{d})^2~.
\end{align}
\noindent The mean-field approximation consists in assuming the cross term $[d_i -\bar{d}][d_j -\bar{d}]$ to sum to zero, i.e.\ that $d_i$ and $d_j$ are uncorrelated. 

For the Ising-like term in (say) the top leaflet, the interaction matrix $V_{uv}$ permits a mapping to the Ising model. Define the exchange parameter $J^\textrm{Ising}$ (unrelated to the $J$ in our model) and the `spin' variable $s_i^\textrm{t} \equiv 2{\hat{\phi}_i^\textrm{t} -1}$, which takes the value $1$ or $-1$, and consider the Ising model in which the interaction energy between spins $i$ and $j$ is $E_{ij} = -J^\textrm{Ising} s_i^\textrm{t} s_j^\textrm{t}$. In the mean-field approximation, the total energy of this Ising model is $E \approx -2 J^\textrm{Ising}(\bar{s}^\textrm{t})^2 N$ where $\bar{s}^\textrm{t}$ is the mean value of the spin. This can be written as
\begin{align}
E = \sum_{<i,j>} -J^\textrm{Ising} s_i^\textrm{t} s_j^\textrm{t} \approx \sum_i - 2 J^\textrm{Ising} s_i^\textrm{t} \bar{s}^\textrm{t}~.
\end{align}

The excess interaction energy for unlike versus like neighbours in the Ising model is $E_{ij}\rvert_{s_i^\textrm{t} = -  s_j^\textrm{t}} - E_{ij}\rvert_{s_i^\textrm{t} =  s_j^\textrm{t}} = 2J^\textrm{Ising}$. For $V_{uv}$, this difference is $V \equiv V_{10} -\tfrac{1}{2}( V_{00}+V_{11})$. Hence equivalence with the Ising model is established by setting $V = 2J^\textrm{Ising}$. Therefore, in the mean-field approximation of our lattice Hamiltonian, we can write
\begin{align}
\sum_{<i,j>} V_{\hat{\phi}_i^\textrm{t} \hat{\phi}_j^\textrm{t}} &\approx \sum_i -V s_i^\textrm{t} \bar{s}^\textrm{t} 
= \sum_i -V s_i^\textrm{t} (2\phi^\textrm{t} - 1)~,
\end{align}
\noindent and similarly for the bottom leaflet. The mean-field (i.e.\ on-site) approximation to the Hamiltonian is thus given by 
\begin{align}
H \approx H_\textrm{MF} = \sum_i H_i~,
\end{align}
\noindent where
\begin{align}
H_i = &-V s_i^\textrm{t} (2\phi^\textrm{t} - 1) -V s_i^\textrm{b} (2\phi^\textrm{b} - 1) 
+ \tfrac{1}{2}J (d_i - \bar{d})^2 
+ \tfrac{1}{2}B (\Delta_i)^2 
+  \tfrac{1}{2}\kappa \left(( \ell^\textrm{t}_i - \ell_0^{\textrm{t}i})^2 + ( \ell^\textrm{b}_i - \ell_0^{\textrm{b}i})^2                      \right)~,
\end{align}
\noindent and $J \equiv 4\tilde{J}$. 

\subsection{Site types}

We now note that, given a mixture of $S$ and $U$ species in each leaflet, there are four possible site types $\alpha \in \{SS,UU,SU,US\}$, where an $AB$ site contains species $A$ on the top and $B$ on the bottom. $SS$ and $UU$ sites are pairwise registered, while $SU$ and $US$ sites are pairwise antiregistered. All sites of type $\alpha$ share the same values of the species-dependent constants $s_i^\textrm{t}$, $s_i^\textrm{b}$, $\ell_0^{\textrm{t}i}$ and $\ell_0^{\textrm{b}i}$ in their $H_i$. We can therefore express the total mean-field Hamiltonian $H_\textrm{MF}$ as a sum over the noninteracting site-level Hamiltonians
\begin{align}
H_\textrm{MF} = \sum_\alpha \sum_{j_\alpha}^{N_\alpha} H_{j_\alpha}~,
\end{align}
\noindent where $j_\alpha$ labels the $j$th out of $N_\alpha$ sites of type $\alpha$, and
\begin{align}\label{eqn:Hjalpha}
H_{j_\alpha} = &\pm V (2 \phi^\textrm{t}-1)  \pm V (2{\phi}^\textrm{b}-1) 
+\tfrac{1}{2}J (d_{j_\alpha} - \bar{d})^2 +\tfrac{1}{2} B (\Delta_{j_\alpha})^2  
+\tfrac{1}{2}\kappa  \left(( \ell^\textrm{t}_{j_\alpha} - \ell_0^{\textrm{t}\alpha})^2 + ( \ell^\textrm{b}_{j_\alpha} - \ell_0^{\textrm{b}\alpha})^2\right)~.
\end{align}
\noindent Here, $\ell_0^{\textrm{t}\alpha} = \ell_{A0}$ and $\ell_0^{\textrm{b}\alpha} = \ell_{B0}$ for $\alpha = AB$, and the $\pm$ are $- -$ for $\alpha = SS$, $+ +$ ($UU$), $- +$ ($SU$) and $+ -$ ($US$).

The sum over the top and bottom leaflet configurations can be rewritten as a sum over the occupancies of the set of site types, i.e.
\begin{align} \label{eqn:Nsum}
\sum^\textrm{constr.}_{\{\hat{\phi}^\textrm{t}_i,\hat{\phi}^\textrm{b}_i\}} = \sum^\textrm{constr.}_{\{N_\alpha\}} \frac{1}{\prod_\alpha N_\alpha!}~,
\end{align}
\noindent where the factorials avoid overcounting indistinguishable configurations and the sum is constrained by Eqs.~\ref{eqn:sc3}--\ref{eqn:sc5}.

Defining $\mathcal{D} \Delta_\alpha \equiv \prod_{j \alpha} d \Delta_{j \alpha}$ and $\mathcal{D} d_\alpha \equiv \prod_{j \alpha} d d_{j \alpha}$, the mean-field partition function $Z_\textrm{MF}$ is
\begin{align}
Z_\textrm{MF} = \sum^\textrm{constr.}_{\{N_\alpha\}} \frac{1}{\prod_\alpha N_\alpha!} \prod_\alpha \int \mathcal{D} \Delta_\alpha \mathcal{D} d_\alpha \exp{(  -\beta \sum_{j_\alpha}^{N_\alpha} H_{j_\alpha}    )}~.
\end{align}
\noindent Since all sites are now independent of one another, the integral may be rewritten in terms of the partition function for a single site of type $\alpha$. Additionally, the constraints Eqs.~\ref{eqn:sc3}--\ref{eqn:sc5} allow the Ising interaction $V$ to be factored out. We thus have
\begin{align}
Z_\textrm{MF} =  \sum^\textrm{constr.}_{\{N_\alpha\}} \exp{\left(-\beta N V^*(\phi^\textrm{t},\phi^\textrm{b})\right)}     \frac{\prod_\alpha Z_\alpha ^{N_\alpha}}{\prod_\alpha N_\alpha!} ~,
\end{align}
\noindent where we have defined 
\begin{align}
V^*(\phi^\textrm{t},\phi^\textrm{b}) \equiv -2V(\phi^\textrm{t}-\phi^\textrm{b})^2-2V(\phi^\textrm{t}+\phi^\textrm{b}-1)^2~.
\end{align}
The single-site thickness partition function is given by
\begin{align}
Z_\alpha = \int d \Delta_{\alpha} d d_{\alpha} \exp{(-\beta H_{\alpha})}~,
\end{align}
\noindent in which
\begin{align} \label{eqn:Hlengthonly}
H_{\alpha} = 
&\tfrac{1}{2}J (d_{\alpha} - \bar{d})^2 +\tfrac{1}{2} B (\Delta_{\alpha})^2 
+\tfrac{1}{2}\kappa  \left(( \ell^\textrm{t}_{\alpha} - \ell_0^{\textrm{t}\alpha})^2 + ( \ell^\textrm{b}_{\alpha} - \ell_0^{\textrm{b}\alpha})^2\right)~,
\end{align}
\noindent now contains only the thickness-dependent interactions. 

\subsection{Self-consistency, free energy}

For self-consistency of the locally averaged bilayer thickness $\bar{d}$ and difference $\overline{\Delta}$, we require the integrations over $d_\alpha,\,\Delta_\alpha$ to be performed subject to Eqs.~\ref{eqn:sc1}, \ref{eqn:sc2}. Since these integrals are Gaussian, and the constraints Eqs.~\ref{eqn:sc1}, \ref{eqn:sc2} are linear, the integrations can be performed exactly to yield
\begin{align}\label{eqn:gaussint}
\prod_\alpha Z_{\alpha}^{N_\alpha} = \exp{(-\beta \sum_\alpha N_\alpha H_\alpha\{d_\alpha^* ,\, \Delta_\alpha^*\})}~,
\end{align}
\noindent where $\{d_\alpha^*,\,\Delta_\alpha^*\}$ minimise $\sum_\alpha N_\alpha H_\alpha$ subject to Lagrange multipliers enforcing Eqs.~\ref{eqn:sc1}, \ref{eqn:sc2}.

Now the mean-field partition function can be written
\begin{widetext}
\begin{align}
Z_\textrm{MF} &=  \sum^\textrm{constr.}_{\{N_\alpha\}} \exp{\left( -\beta \left[ N V^*(\phi^\textrm{t},\phi^\textrm{b})  + \sum_\alpha N_\alpha \left(H_\alpha\{d_\alpha^* ,\, \Delta_\alpha^*\} + k_\textrm{B}T \ln{N_\alpha}\right)                              \right]                \right)}\notag\\
 &\equiv \sum^\textrm{constr.}_{\{N_\alpha\}} \exp{(-\beta N \tilde{f})}~,
\end{align}
\end{widetext}
where Stirling's approximation has been used ($\ln{N_\alpha!} \approx N_\alpha \ln{N_\alpha} - N_\alpha$), contributing an irrelevant constant.

The three constraints Eqs.~\ref{eqn:sc3}--\ref{eqn:sc5} leave only one $N_\alpha$ over which to sum. For this we perform a saddle-point approximation, which is equivalent to removing the sum and setting $\{N_\alpha\}$ to their values that minimise $\tilde{f}$ subject to Eqs.~\ref{eqn:sc3}--\ref{eqn:sc5}. This yields 
\begin{align}
Z_\textrm{MF} \approx \exp{(-\beta N \tilde{f}^*)}~,
\end{align}
\noindent where $\tilde{f}^*$ is the minimised value of $\tilde{f}$. Then, by Eq.~\ref{eqn:bythis}, our desired free energy per site $f(\phi^\textrm{t},\,\phi^\textrm{b},\,\bar{d},\,\overline{\Delta})$ is given by $\tilde{f}^*$.

The steps described above can be summarised compactly by stating that the free energy $f(\phi^\textrm{t},\,\phi^\textrm{b},\,\bar{d},\,\overline{\Delta})$ per site is given by minimising 
\begin{widetext}
\begin{align} \label{eqn:MFF2}
f^{'} N = &\sum_\alpha N_\alpha \left(H_{\alpha} + k_\textrm{B}T \ln{N_\alpha}\right) 
- 2 V N(\phi^\textrm{t} - \phi^\textrm{b})^2 -  2 V N(\phi^\textrm{t} + \phi^\textrm{b} -1)^2~
\end{align}
over $\{d_\alpha,\,\Delta_\alpha,\,N_\alpha$\} subject to Lagrange multipliers enforcing Eqs.~\ref{eqn:sc1}--\ref{eqn:sc5}, as written in Eq.~\ref{eqn:MFF}.
The variables fixed in the minimisation procedure are 
\end{widetext}
\begin{subequations} \label{eqn:ingredients1}
\begin{align} 
d_{SS} &= \bar{d} + \frac{\kappa \Delta_0}{2 J + \kappa} (2 - \phi^\textrm{t} - \phi^\textrm{b} )~,\\
d_{UU} &= \bar{d}- \frac{\kappa \Delta_0}{2 J + \kappa} ( \phi^\textrm{t} + \phi^\textrm{b})~,\\
d_{SU} &= d_{US} =  \bar{d}- \frac{\kappa \Delta_0}{2 J + \kappa} ( \phi^\textrm{t} + \phi^\textrm{b}-1)~,\\
\Delta_{SS} &=  \Delta_{UU} = \overline{\Delta}-  \frac{\kappa \Delta _0}{2 B + \kappa} (\phi^\textrm{t}-\phi^\textrm{b})~,\\
\Delta_{SU} &=  \overline{\Delta}-  \frac{\kappa \Delta_0}{2 B + \kappa} (\phi^\textrm{t}-\phi^\textrm{b}-1)~,\\
\Delta_{US} &=  \overline{\Delta}-  \frac{\kappa \Delta_0}{2 B + \kappa} (\phi^\textrm{t}-\phi^\textrm{b}+1)~,\\
N_{SS}/N &= A(\phi^\textrm{t}, \phi^\textrm{b})~,\\
N_{UU}/N &= A(\phi^\textrm{t}, \phi^\textrm{b})+ 1 - \phi^\textrm{t } - \phi^\textrm{b} ~,\\
N_{SU}/N &= -A(\phi^\textrm{t}, \phi^\textrm{b})+\phi^\textrm{t}~,\\
N_{US}/N &= -A(\phi^\textrm{t}, \phi^\textrm{b})+\phi^\textrm{b}~.
\end{align}
\end{subequations}

We have defined
\begin{align} \label{eqn:ingredients2}
A(\phi^\textrm{t}, \phi^\textrm{b}) \equiv \frac{2\phi^\textrm{t}\phi^\textrm{b}}
{\phi^*+\sqrt{\phi^{*\,2} + 4\phi^\textrm{t}\phi^\textrm{b}(e^{-2\beta\sigma}-1)}}~,
\end{align}
\noindent where
\begin{align}
\phi^* \equiv \phi^\textrm{t}+ \phi^\textrm{b} +  e^{-2\beta\sigma}     (1-\phi^\textrm{t} -\phi^\textrm{b})~,
\end{align}
\noindent and
\begin{align}\label{eqn:Xappendix}
\sigma &\equiv \tfrac{1}{2}(H_{SU} + H_{US} - H_{SS} -H_{UU}) \notag \\
&= - \frac{\Delta_0^2 \kappa^2 (J-B)}{2(2 J + \kappa) (2 B + \kappa)}~, 
\end{align}
\noindent is the energy change per site for converting two R sites into two AR sites. The expected self-consistency requirements are fulfilled;\ for example, $\phi^\textrm{b} = \phi^\textrm{t} \to 1$ (forcing all sites to be of $SS$ type) leads to $d_{SS} \to \bar{d}$. 

To construct the local free energy $f(\phi^\textrm{t},\,\phi^\textrm{b},\,\bar{d},\,\overline{\Delta})$, we insert Eqs.~\ref{eqn:ingredients1}--\ref{eqn:Xappendix} into Eq.~\ref{eqn:MFF2}. We find
\begin{widetext}
\begin{align} \label{eqn:fullf}
&f(\phi^\textrm{t},\,\phi^\textrm{b},\,\bar{d},\,\overline{\Delta}) = \notag \\
&k_\textrm{B} T \left[ A\ln{A} + (A+1-\phi^\textrm{t} - \phi^\textrm{b}) \ln{(A+1 - \phi^\textrm{t} - \phi^\textrm{b})} + (\phi^\textrm{t}-A) \ln{(\phi^\textrm{t}-A)} + (\phi^\textrm{b} -A) \ln{(\phi^\textrm{b}-A)} \right] \notag \\
&+\frac{1}{2} \kappa \left[ {\frac{1}{2}}(\bar{d} - d_0)^2 + \Delta_0 \left( (\phi^\textrm{t} +\phi^\textrm{b} -1)(d_0 -\bar{d})-(\phi^\textrm{t}-\phi^\textrm{b}) \overline{\Delta} + \frac{1}{2} \Delta_0 \right) \right] + \frac{1}{4} \overline{\Delta}^2 (2B + \kappa) \notag \\
&+\frac{\kappa^2 \Delta_0^2}{2(2B+\kappa)(2J+\kappa)} \left[\left(2A - 2 \phi^\textrm{t} \phi^\textrm{b}\right)\left(J-B\right) - \left(\phi^\textrm{t} + \phi^\textrm{b} - {\phi^{\textrm{t}\,2}} - {\phi^{\textrm{b}\,2}}\right)\left({J+B+\kappa} \right)\right] \notag \\
&- 2 V (\phi^\textrm{t} - \phi^\textrm{b})^2 
-  2 V (\phi^\textrm{t} + \phi^\textrm{b} -1)^2~,
\end{align}
\end{widetext}
\noindent where $\Delta_0 \equiv \ell_{S0} - \ell_{U0}$, $d_0 \equiv \ell_{S0} + \ell_{U0}$. 

Upon further minimising $f$ over the mean-field thickness variables $\bar{d}$ and $\overline{\Delta}$, we obtain $f^\textrm{[ann.]}(\phi^\textrm{t},\phi^\textrm{b})$, which determines the minima in the local free energy landscape (see e.g.\ Figs.~\ref{botleft}, \ref{surfaces}). The annealed thickness variables are 
\begin{subequations}
\begin{align}
\bar{d}^\textrm{[ann.]} &= \Delta_0 (\phi^\textrm{t} + \phi^\textrm{b} -1) + d_0~, \\
\overline{\Delta}^\textrm{[ann.]} &= \frac{\kappa \Delta_0 (\phi^\textrm{t} - \phi^\textrm{b})}{2 B + \kappa}~,
\end{align}
\end{subequations}
\noindent giving
\begin{widetext}
\begin{align} \label{eqn:fullfann}
&f^\textrm{[ann.]}(\phi^\textrm{t},\phi^\textrm{b}) =  \notag \\
&k_\textrm{B} T \left[ A\ln{A} + (A+1-\phi^\textrm{t} - \phi^\textrm{b}) \ln{(A+1 - \phi^\textrm{t} - \phi^\textrm{b})} + (\phi^\textrm{t}-A) \ln{(\phi^\textrm{t}-A)} + (\phi^\textrm{b} -A) \ln{(\phi^\textrm{b}-A)} \right] \notag \\
&+\frac{1}{2}\frac{B\kappa\Delta_0^2(\phi^\textrm{t}+\phi^\textrm{b})}{2B+\kappa}\left(2-\phi^\textrm{t}-\phi^\textrm{b}\right)-\sigma\left(2A+[\phi^\textrm{t}+\phi^\textrm{b}][1-\phi^\textrm{t}-\phi^\textrm{b}]\right)\notag\\
&- 2 V (\phi^\textrm{t} - \phi^\textrm{b})^2 
-  2 V (\phi^\textrm{t} + \phi^\textrm{b} -1)^2~.
\end{align}
\end{widetext}

\section{\label{app:ginzburg}Ginzburg-Landau analysis}

$f (\phi^\textrm{t},\,\phi^\textrm{b},\,\bar{d},\,\overline{\Delta})$ is the coarse-grained free energy per site. This can serve as the `Landau part' of a Ginzburg-Landau type free energy $F_\textrm{G-L}$ to study kinetics
\begin{align} \label{eqn:SIginzburg}
F_\textrm{G-L} = \int d^2 r \left(\frac{f}{a^2} + f_\textrm{grad} \right)~,
\end{align}
\noindent where
\begin{align}
f_\textrm{grad} = \frac{\tilde{J}}{2}(\nabla \bar{d})^2 + {V}(\nabla \phi^\textrm{t})^2 + {V} (\nabla \phi^\textrm{b})^2~.
\end{align}

\noindent This gradient contribution depends on the terms of the Hamiltonian by which laterally neighbouring sites interact. The composition gradient term in each leaflet involving $V$ is simply that for the mean-field Ising model \cite{Goldenfeld}, and the thickness gradient term involving $\tilde{J}$ is the corresponding term of Eq.~\ref{eqn:fun4} in the limit of small lattice spacing.

We study the instabilities about a reference `homogeneous state' defined by $\phi^\textrm{b}\!=\!\phi^\textrm{t}\!=\!0.5$, $\overline{\Delta}= 0$, $\bar{d}= d_0$, applying small perturbations to this state and determining the resultant change in $F_\textrm{G-L}$. The thermodynamic driving force for instability to demixing, determined by $f$, competes with the gradient terms $f_\textrm{grad}$ which penalise the resultant inhomogeneity. Combining these with evolution equations for composition and thickness, we find preferred lengthscales for initial demixing, and associated rates, that can be compared between the R and AR modes.

A perturbation is described by $\delta \phi^\textrm{t},\,\delta \phi^\textrm{b},\,\delta \bar{\ell}^\textrm{t},\, \delta \bar{\ell}^\textrm{b}$. Considering separately the R mode (in which $\phi^\textrm{b}_\textrm{R}(\phi^\textrm{t}) = \phi^\textrm{t}$, $\overline{\Delta}= 0$ and $\delta \bar{\ell}^\textrm{b} = \delta \bar{\ell}^\textrm{t}$) and the AR mode (in which $\phi^\textrm{b}_\textrm{AR}(\phi^\textrm{t}) = 1-\phi^\textrm{t}$, $\bar{d}= d_0$ and $\delta \bar{\ell}^\textrm{b} =- \delta \bar{\ell}^\textrm{t}$), we now apply linear stability analysis to perturbations governed by Eq.~\ref{eqn:SIginzburg}, to determine which mode initially grows fastest.

\begin{widetext}
\subsection{Evolution of perturbations}
The free energy change due to a perturbation in mode $\textrm{m}$ is
\begin{align}
\begin{aligned}
\delta F_\textrm{G-L}^\textrm{m} = \frac{1}{2}\int d^2r \Bigg(\begin{pmatrix}\delta \phi^\textrm{t} \\ \delta \bar{\ell}^\textrm{t} \end{pmatrix} 
 \cdot  \underline{\underline{C}}^\textrm{m} \cdot \begin{pmatrix}\delta \phi^\textrm{t} \\ \delta \bar{\ell}^\textrm{t} \end{pmatrix}   
 + \nabla \begin{pmatrix}\delta \phi^\textrm{t} \\ \delta \bar{\ell}^\textrm{t} \end{pmatrix}  \cdot \underline{\underline{P}}^\textrm{m} \cdot \nabla\begin{pmatrix}\delta \phi^\textrm{t} \\ \delta \bar{\ell}^\textrm{t} \end{pmatrix}  \Bigg)~,
\end{aligned}
\end{align}
\noindent where $\textrm{m} = \textrm{R, AR}$ and the matrices $\underline{\underline{C}}^\textrm{m}$ and $ \underline{\underline{P}}^\textrm{m}$ respectively contain the bulk and gradient free energy terms:

\begin{subequations}
\begin{align}
a^2 \underline{\underline{C}}^\textrm{R} &= 
\begin{bmatrix}  
 f^\textrm{R}_{\phi^\textrm{t}\phi^\textrm{t}}  &  2 f^\textrm{R}_{\bar{d}\phi^\textrm{t}}  \\[10pt]
 2 f^\textrm{R}_{\bar{d}\phi^\textrm{t}}  &  4 f^\textrm{R}_{\bar{d}\bar{d}} 
\end{bmatrix}
=
2\begin{bmatrix}  
\Bigg( {2k_\textrm{B}T\left[e^{-\beta\sigma} +1\right]} 
+ \frac{\kappa^2 \Delta_0^2}{2J+\kappa}- 
8 V\Bigg) & - \kappa \Delta_0  \\[10pt]
- \kappa \Delta_0 &   \kappa
\end{bmatrix}~, \\
a^2\underline{\underline{C}}^\textrm{AR} &= 
\begin{bmatrix}  
  f^\textrm{AR}_{\phi^\textrm{t}\phi^\textrm{t}} & 2 f^\textrm{AR}_{ {\scriptscriptstyle \overline{\Delta}} \phi^\textrm{t}}  \\[10pt]
2  f^\textrm{AR}_{ {\scriptscriptstyle \overline{\Delta}} \phi^\textrm{t}}  &   4 f^\textrm{AR}_{ \scriptscriptstyle \overline{\Delta} \, \overline{\Delta}} 
\end{bmatrix}
=
2\begin{bmatrix}  
 \Bigg({2k_\textrm{B}T\left[              e^{\beta\sigma} + 1\right]}
+\frac{\kappa^2 \Delta_0^2}{2B+\kappa} -
 8 V\Bigg)& -\kappa \Delta_0  \\[10pt]
-\kappa \Delta_0&    2B+\kappa
\end{bmatrix}~, \\
\underline{\underline{P}}^\textrm{R} &= 
\begin{bmatrix}
4V & 0 \\[10pt]
0 & 4\tilde{J}
\end{bmatrix}~,~
 \underline{\underline{P}}^\textrm{AR} = 
\begin{bmatrix}
4V & 0 \\[10pt]
0 & 0
\end{bmatrix}~.
\end{align}
\end{subequations}
\noindent $f^\textrm{R}(\phi^\textrm{t},\,\bar{d})$ represents $f$ evaluated for $\phi^\textrm{b}_\textrm{R}(\phi^\textrm{t}) = \phi^\textrm{t}$, $\overline{\Delta}= 0$, and $f^\textrm{AR}(\phi^\textrm{t},\,\overline{\Delta})$ represents $f$ evaluated for $\phi^\textrm{b}_\textrm{AR}(\phi^\textrm{t}) = 1-\phi^\textrm{t}$, $\bar{d}= d_0$. Subscripts indicate derivatives evaluated at the homogeneous state, i.e.\ $\phi^\textrm{t} = 0.5$, $\bar{d}= d_0$, $\overline{\Delta}= 0$.

\end{widetext}

Since composition is conserved, it evolves \cite{Wallace2005} via 
\begin{align}
\frac{\partial \delta \phi^\textrm{t m}}{\partial t} = &M \nabla^2 ( C_{11}^\textrm{m} \delta \phi^\textrm{t} + C_{12}^\textrm{m}  \delta \bar{\ell}^\textrm{t} - P_{11}^\textrm{m} \nabla^2 \delta \phi^\textrm{t} - P_{12}^\textrm{m} \nabla^2  \delta \bar{\ell}^\textrm{t} )~,
\end{align}
\noindent where the mobility $M$ sets the timescale. 

We assume thickness to behave in a nonconserved fashion so that it evolves relaxationally \cite{Wallace2005}, via 
\begin{align}
\frac{\partial \delta \bar{\ell}^\textrm{t m}}{\partial t} = &-\eta ( C_{21}^\textrm{m}  \delta \phi^\textrm{t} + C_{22}^\textrm{m}  \delta \bar{\ell}^\textrm{t} - P_{21}^\textrm{m} \nabla^2 \delta \phi^\textrm{t} - P_{22}^\textrm{m} \nabla^2  \delta \bar{\ell}^\textrm{t} )~,
\end{align}
\noindent where the mobility $\eta$ incorporates frictional forces involved in length stretching and compression of amphiphiles (in principle it can acquire wavenumber dependence via coupling to the conserved solvent flow). 

In Fourier space, the coupled evolution equations are
\begin{align}
\frac{\partial}{\partial t} \begin{pmatrix}\delta \phi^\textrm{t m}_q \\ \delta \bar{\ell}^\textrm{t m}_q \end{pmatrix} &= - \underline{\underline{M}}(q) \cdot (\underline{\underline{C}}^\textrm{m} + q^2 \underline{\underline{P}}^\textrm{m}) \cdot \begin{pmatrix}\delta \phi^\textrm{t}_q \\ \delta \bar{\ell}^\textrm{t}_q \end{pmatrix}\ \notag \\
&\equiv -\underline{\underline{L}}^\textrm{m}(q) \cdot \begin{pmatrix}\delta \phi^\textrm{t}_q \\ \delta \bar{\ell}^\textrm{t}_q \end{pmatrix}~,
\end{align}
\noindent where
\begin{align}
\underline{\underline{M}}(q) \equiv
\begin{pmatrix}
 Mq^2 & 0 \\ 0 & M \xi
\end{pmatrix}~.
\end{align}
\noindent The dimensionless parameter $\xi \equiv \eta / M$ controls how `fast' the thickness relaxation is relative to diffusion. Instabilities of the R or AR mode correspond to a negative eigenvalue of their $\underline{\underline{L}}^\textrm{m}$. Their wavenumber dependent growth rates are given by $\omega^\textrm{m}(q) = - \lambda^\textrm{m}$ where $\lambda^\textrm{m}$ is the eigenvalue for the eigenmode of $\underline{\underline{L}}^\textrm{m}$. Maximising $\omega^\textrm{m}(q)$ over $q$ yields $\omega^\textrm{m}_\textrm{max}$, the peak growth rate of the given mode (R or AR). 

The blue and red colours in Fig.~\ref{JBlines} are obtained by first calculating $\omega^\textrm{m}_\textrm{max}$ for the R and AR modes. Then the difference $\Delta \omega \equiv \omega^\textrm{R}_\textrm{max} - \omega^\textrm{AR}_\textrm{max}$ is plotted as the background of Fig.~\ref{JBlines}. If a given mode $\textrm{m}$ has a negative peak growth rate (i.e.\ is not unstable) then its $\omega^\textrm{m}_\textrm{max}$ is set to zero. Thus $\Delta \omega = 0$ (white) is ambiguous;\ either i)~both modes are stable or ii)~both are unstable but with equal peak growth rates. This ambiguity is easily resolved by referring to the instability lines when interpreting the plot, since if mode $\textrm{m}$'s peak growth rate is zero then we must be outside the instability region of mode $\textrm{m}$. Note that the ranges of the colour scales in Fig.~\ref{JBlines} are asymmetric. 

\begin{figure}[floatfix]
\includegraphics[width=8.25cm]{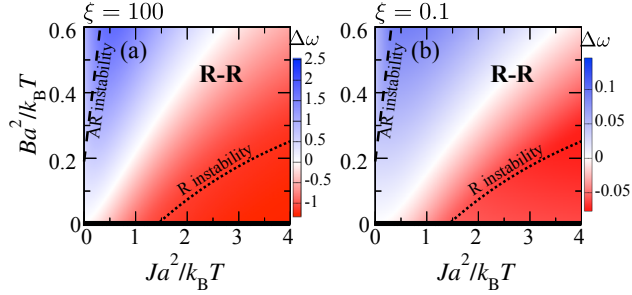}
\caption{\label{xis}(a)~As Fig.~\ref{JBlines}d. (b)~With $\xi=0.1$ instead of $\xi=100$.} 
\end{figure}

To model the physically likely scenario we have used $\xi = 100$, since any frictional drag involved in stretching should be far less than that for lateral diffusion \cite{Wallace2005, Wallace2006}. This value is close to `saturation', i.e.\ the composition relaxation is the limiting timescale and significant further increases in $\xi$ have only marginal quantitative effects on $\omega^\textrm{m}(q)$. 
Therefore the conclusions drawn from the colours in Fig.~\ref{JBlines} are independent of $\xi$ in the expected physical regime. Even if the opposite regime is assumed ($\xi = 0.1$), the values of $\omega^\textrm{m}(q)$ change but the key feature of the $\Delta \omega$  landscape -- which mode is fastest -- is not strongly affected (Fig.~\ref{xis}). 

\section{\label{app:physical}Physical meaning of parameters}

For comparison with phospholipids, we set the lattice spacing $a \sim 0.8\,\textrm{nm}$, corresponding to an area per lipid of $0.64\,\textrm{nm}^2$ \cite{Kucerka2011}. 

\subsection{Stretching modulus}
The stretching and compression of a bilayer of amphiphiles is measured experimentally via the area stretching modulus $\kappa_A$, with a free energy given by
 \begin{equation}
 G_{\kappa_A}^\textrm{bilayer} = \int d^2 r \frac{\kappa_A}{2}\left(\frac{\delta\! A}{A_0}\right)^2~,
\label{eqn:kappaarea}
\end{equation}
\noindent where $\delta\! A$ represents an area difference relative to the equilibrium area $A_0$. In the continuum representation of the lattice model here the stretching free energy for an individual leaflet of the bilayer is given by
 \begin{equation}
G_{\kappa}^\textrm{leaflet} = \int \frac{d^2 r}{a^2} \frac{{\kappa} }{2}\left(\delta \ell \right)^2~,
\label{eqn:kappalength}
\end{equation}
\noindent where $a^2$ is the lattice site area and $\delta \ell$ is a tail length difference relative to an equilibrium length $\ell_0$. Assuming that the volume $v = A\ell$ remains constant upon stretching/compression, $A \delta \ell = - \ell \delta\! A$, we have
 \begin{equation}
G_{\kappa}^\textrm{leaflet}   =  \int d^2 r \frac{ \ell_0^2}{a^2} \frac{{\kappa} }{2}\left(\frac{\delta\! A}{A_0}\right)^2~.
\label{eqn:kappalength2}
\end{equation}
\noindent Noting that Eq.~\ref{eqn:kappaarea} describes the area stretching energy for the \textit{whole} bilayer, and assuming the energy to be distributed evenly between the two leaflets, we can write
\begin{equation}
G_{\kappa}^\textrm{leaflet}  = \tfrac{1}{2}  G_{\kappa_A}^\textrm{bilayer}~.
\end{equation}
\noindent Identifying the lattice site area $a^2$ as the equilibrium area per amphiphile $A_0$ gives the correspondence
\begin{equation}
\label{eqn:arealength}
{\kappa}  = \frac{A_0}{2\ell_0^2} \kappa_{A}~,
\end{equation}
 \noindent where $\ell_0$ is a representative value for the equilibrium length of a real amphiphile. For typical values $\ell_0 = 2\,\textrm{nm}$ and $A_0 = 0.64\,\textrm{nm}^2$ for phospholipid leaflets, the value $\kappa = 3 \,a^{-2}k_\textrm{B}T$ used in this work corresponds to $\kappa_{A} \approx 40\,a^{-2}k_\textrm{B}T \approx 60\,k_\textrm{B}T\textrm{nm}^{-2}$, in the range for lipid bilayers at $300\,\textrm{K}$ \cite{Wallace2005, Needham1990, Rawicz2000}.
 
 \subsection{Indirect coupling} 
The indirect coupling parameter $\tilde{J}$ quantifies the penalty for mismatch in the total hydrophobic thickness between neighbouring lattice sites, arising from hydrophobic surface tension.
We take a fiducial value $\tilde{J} \approx 0.8\,k_\textrm{B}T\textrm{nm}^{-2}$, approximately that estimated in \cite{Wallace2006} as a surface tension for hydrocarbon tails in contact with the watery headgroup region of phospholipids. This gives $\tilde{J} \approx 0.5\,a^{-2}k_\textrm{B}T$ for the lattice model, so for the mean-field parameter $J \approx 2\,a^{-2}k_\textrm{B}T$. Increasing $J$ (Fig.~\ref{JBlines}) can be thought of as increasing hydrophobic mismatch/hydrophobicity. Our model cannot capture all the intricacies of hydrophobic mismatch between molecules. Instead, through $J$ and its interplay with $\kappa$, we mean to capture the fact that such mismatch is disfavoured \textit{at the molecular scale} \cite{Perlmutter2011, Stevens2005, Zhang2004, *Zhang2007}, and to estimate a reasonable scale for the energy penalty involved.

\subsection{Direct coupling}

The direct coupling parameter $B$ plays a similar role to the inter-leaflet mismatch free energy $\gamma$ estimated in the literature. We can define an effective $\gamma$ (that shown in Fig.~\ref{JBlines}) by considering an isolated AR site and minimising its energy arising from stretching ($\kappa$) and direct coupling ($B$) energies over the top and bottom amphiphile lengths, where the reference state is an isolated R site which experiences zero direct coupling energy.  This microscopic energy per AR site is
 \begin{equation}
\gamma a^2 = \frac{\Delta_0^2 \kappa B}{2 (\kappa+2B)}~,
\label{eqn:mismatchenergy}
\end{equation}
\noindent in terms of which
\begin{equation}
B = \frac{2 \gamma a^2 \kappa}{\kappa \Delta_0^2 - 4 \gamma a^2}~.
\end{equation}
\noindent For example, the value $\gamma \approx 0.15\,k_\textrm{B}T \textrm{nm}^{-2}$ estimated in \cite{Risselada2008, *Polley2013} is, in model units, $\gamma \approx 0.1\,a^{-2}k_\textrm{B}T$. Assuming $\Delta_0 = 1\,a$ and $\kappa=3 \,a^{-2}k_\textrm{B}T$, this gives $B \approx 0.23\,a^{-2}k_\textrm{B}T$. However, even compared to the other parameters of our idealised model, $\gamma$ is poorly understood. Ref.~\cite{Putzel2011} estimates an order of magnitude lower ($\gamma \sim 0.01\,a^{-2}k_\textrm{B}T$, so that $B \sim 0.02\,a^{-2}k_\textrm{B}T$), and finds that the method used to extract $\gamma$ in simulation \cite{Risselada2008, *Polley2013} is inaccurate, since it assumes larger characteristic fluctuations than were measured. On the other hand, \cite{Pantano2011} finds that the effective $\gamma$ measured while artificially pulling domains out of registration depends strongly on mismatch area, and proposes a role for membrane curvature, which we have not studied.

\subsection{Interpretation of $\gamma$} 

There are subtleties in defining the mismatch free energy per area $\gamma$. We have defined it `microscopically' in Eq.~\ref{eqn:mismatchenergy} as the direct coupling energy density for an antiregistered site. It is possible instead to construct a `macroscopic' definition by comparing the free energies of antiregistered and registered \textit{domains} 
\begin{align} \label{eqn:gammamacro}
\gamma_\textrm{macro} \equiv \lim_{A\to \infty} \frac{1}{A}(G^\textrm{antireg}(A) - G^\textrm{reg}(A))~,
\end{align}
\noindent where $G^\textrm{(anti)reg}(A)$ is the free energy of an (anti)registered domain of area $A$. The limit $A \to \infty$ emphasises that boundary contributions to the free energies are typically ignored \cite{Putzel2011, May2009}. For example, in \cite{Putzel2011}, $\gamma_\textrm{macro}$ is computed theoretically by comparing the free energies of antiregistered and registered arrangements of domains within a molecular mean-field theory, the domains being assumed large enough that contributions from their boundaries can be neglected. It is important to note that any effects of hydrophobic mismatch energy at the edges of registered domains (incorporated in our Ginzburg-Landau analysis via the $\tilde{J}$ term of $f_\textrm{grad}$) cannot be properly captured by $\gamma$ or $\gamma_\textrm{macro}$, since these describe only energies that scale as the domain area.

In the well-segregated limit such that an anti(registered) domain contains \textit{purely} pairwise (anti)registered sites, the definition Eq.~\ref{eqn:gammamacro} becomes equivalent to Eq.~\ref{eqn:mismatchenergy}. \textit{Near} this limit, within our model $\gamma_\textrm{macro} \approx \gamma$, because the dominant contribution to the free energy difference in Eq.~\ref{eqn:gammamacro} will be from the direct coupling energy experienced by AR sites (Eq.~\ref{eqn:mismatchenergy}), while contributions associated with the remnant fraction of pairwise R sites in the antiregistered demixed phase (and vice versa) will be small. Thus, for example, the free energy difference between the R and AR minima of $f^\textrm{[ann.]}$ (Fig.~\ref{botleft}) is similar to the value of $\gamma$ quoted for that parameter point on Fig.~\ref{JBlines}e, calculated with Eq.~\ref{eqn:mismatchenergy}.

In general, however, Eq.~\ref{eqn:gammamacro} requires specification of the compositions of the R and AR phases whose free energies are to be compared, and Fig.~\ref{surfaces} shows us that the leaflet compositions in the AR phases generally differ from the those in the R phases. Therefore, the assumption \cite{Putzel2011} that the relevant AR configuration for comparison is that obtained by re-arranging the domains from the R configuration, \textit{without altering their compositions}, is incorrect. It may be suitable for describing small fluctuations into AR at the boundary of a large R domain (as was the purpose in \cite{Putzel2011}), but only if one assumes that spatial fluctuations of the domain boundaries out of registration are not also accompanied by compositional fluctuations of the domains in each leaflet. 
 
In some situations, the relationship between $\gamma$ and $\gamma_\textrm{macro}$ is complicated by ambiguity in implementing the macroscopic definition. Given registered domain coexistence, one might assume that we should take a metastable AR rearrangement of the domains for comparison. However, in Fig.~\ref{surfaces} (top pane), no AR minima exist in the free energy, so moving R-R coexisting domains into antiregistration \cite{Putzel2011} would not yield a metastable state. A single AR phase could still exist as part of R-R-AR equilibrium, depending on the free energy's detailed shape, but the intuitive `R-R to AR-AR' rearrangement used in \cite{Putzel2011} becomes difficult to interpret. 
In another case, in a small region of Fig.~\ref{JBlines}b the AR minima are lower in free energy, so an AR-AR or AR-AR-R state becomes \textit{equilibrium}. Under Eq.~\ref{eqn:gammamacro} this would imply a \textit{negative} value of $\gamma_\textrm{macro}$, although the per-site $\gamma$ defined by Eq.~\ref{eqn:mismatchenergy} is positive. 

Hence, it is clear that describing inter-leaflet coupling is complex, both in terms of specifying the relevant bulk free energy and in terms of the domain size-dependent competition of edge and area energies. This latter aspect in particular, and its role in nucleation kinetics of domain registration, will be further studied in future work. In relation to the present discussion, it is unclear precisely which coupling or combination of couplings is being measured in molecular simulation studies of inter-leaflet coupling \cite{Risselada2008, *Polley2013}, where the probability of fluctuations into antiregistration is monitored and fit to a Boltzmann distribution. These fluctuations may be subject to effects related to hydrophobic mismatch and composition dependence as discussed above, so that even if the approach of measuring fluctuations is essentially correct (challenged in \cite{Putzel2011}), it is likely that energies additional to that described by $\gamma$ are at work. In summary, much further work is required in defining, measuring, and studying the implications of the competing forms of inter-leaflet coupling.

\textbf{Author contributions:}\ JJW designed the model, performed the research, and wrote the paper. PDO designed the model, supervised the research and wrote the paper. 

\begin{acknowledgments} 
We acknowledge the EPSRC CAPITALS grant (EP/J017566/1) and discussions with A Aufderhorst-Roberts, HMG Barriga, NJ Brooks, P Cicuta, SD Connell, E Del Gado, SL Keller, H Kusumaatmaja, N McCarthy, D Rings, JM Seddon and AP Tabatabai. The input of anonymous referees is greatly appreciated.

This work was initiated at the University of Leeds, Leeds, United Kingdom, and was funded by EPSRC and Georgetown University. PDO gratefully acknowledges the support of the Ives endowment.
\end{acknowledgments} 

\bibliography{bibliography}

\begin{thebibliography}{49}
\providecommand{\url}[1]{\texttt{#1}}
\providecommand{\urlprefix}{ }

\bibitem[Lingwood and Simons(2010)]{Lingwood2010}
Lingwood, D., and K.~Simons, 2010.
\newblock Lipid Rafts As a Membrane-Organizing Principle.
\newblock \emph{Science} 327:46--50.

\bibitem[Kusumi et~al.(2004)Kusumi, Koyama-Honda, and Suzuki]{Kusumi2004}
Kusumi, A., I.~Koyama-Honda, and K.~Suzuki, 2004.
\newblock Molecular Dynamics and Interactions for Creation of
  Stimulation-Induced Stabilized Rafts from Small Unstable Steady-State Rafts.
\newblock \emph{Traffic} 5:213--230.

\bibitem[May(2009)]{May2009}
May, S., 2009.
\newblock Trans-monolayer coupling of fluid domains in lipid bilayers.
\newblock \emph{Soft Matter} 5:3148--3156.

\bibitem[Garb{\`e}s~Putzel et~al.(2011)Garb{\`e}s~Putzel, Uline, Szleifer, and
  Schick]{Putzel2011}
Garb{\`e}s~Putzel, G., M.~J. Uline, I.~Szleifer, and M.~Schick, 2011.
\newblock Interleaflet Coupling and Domain Registry in Phase-Separated Lipid
  Bilayers.
\newblock \emph{Biophys. J.} 100:996--1004.

\bibitem[Funkhouser et~al.(2013)Funkhouser, Mayer, Solis, and
  Thornton]{Funkhouser2013}
Funkhouser, C.~M., M.~Mayer, F.~J. Solis, and K.~Thornton, 2013.
\newblock Effects of interleaflet coupling on the morphologies of
  multicomponent lipid bilayer membranes.
\newblock \emph{J. Chem. Phys.} 138:024909.

\bibitem[Allender and Schick(2006)]{Allender2006}
Allender, D., and M.~Schick, 2006.
\newblock Phase Separation in Bilayer Lipid Membranes: Effects on the Inner
  Leaf Due to Coupling to the Outer Leaf.
\newblock \emph{Biophys. J.} 91:2928--2935.

\bibitem[Garb{\`e}s~Putzel and Schick(2008)]{Putzel2008}
Garb{\`e}s~Putzel, G., and M.~Schick, 2008.
\newblock Phase Behavior of a Model Bilayer Membrane with Coupled Leaves.
\newblock \emph{Biophys. J.} 94:869--877.

\bibitem[Wagner et~al.(2007)Wagner, Loew, and May]{Wagner2007}
Wagner, A.~J., S.~Loew, and S.~May, 2007.
\newblock Influence of Monolayer-Monolayer Coupling on the Phase Behavior of a
  Fluid Lipid Bilayer.
\newblock \emph{Biophys. J.} 93:4268--4277.

\bibitem[Hirose et~al.(2009)Hirose, Komura, and Andelman]{Hirose2009}
Hirose, Y., S.~Komura, and D.~Andelman, 2009.
\newblock Coupled Modulated Bilayers: A Phenomenological Model.
\newblock \emph{ChemPhysChem} 10:2839--2846.

\bibitem[Korlach et~al.(1999)Korlach, Schwille, Webb, and
  Feigenson]{Korlach1999}
Korlach, J., P.~Schwille, W.~W. Webb, and G.~W. Feigenson, 1999.
\newblock Characterization of lipid bilayer phases by confocal microscopy and
  fluorescence correlation spectroscopy.
\newblock \emph{Proc. Natl. Acad. Sci.} 96:8461--8466.

\bibitem[Dietrich et~al.(2001)Dietrich, Bagatolli, Volovyk, Thompson, Levi,
  Jacobson, and Gratton]{Dietrich2001}
Dietrich, C., L.~Bagatolli, Z.~Volovyk, N.~Thompson, M.~Levi, K.~Jacobson, and
  E.~Gratton, 2001.
\newblock Lipid Rafts Reconstituted in Model Membranes.
\newblock \emph{Biophys. J.} 80:1417--1428.

\bibitem[Collins and Keller(2008)]{Collins2008}
Collins, M.~D., and S.~L. Keller, 2008.
\newblock Tuning lipid mixtures to induce or suppress domain formation across
  leaflets of unsupported asymmetric bilayers.
\newblock \emph{Proc. Natl. Acad. Sci.} 105:124--128.

\bibitem[Note1()]{Note1}
The mismatch free energy per area can be defined as $\gamma \equiv (G^\protect
  \textrm {antireg}(A) - G^\protect \textrm {reg}(A))/A$, where $G^\protect
  \textrm {(anti)reg}(A)$ is the free energy of a large (anti)registered domain
  of area $A$.

\bibitem[Garc\'{i}a-S\'{a}ez et~al.(2007)Garc\'{i}a-S\'{a}ez, Chiantia, and
  Schwille]{Garcia2007}
Garc\'{i}a-S\'{a}ez, A.~J., S.~Chiantia, and P.~Schwille, 2007.
\newblock Effect of Line Tension on the Lateral Organization of Lipid
  Membranes.
\newblock \emph{J. Biol. Chem.} 282:33537--33544.

\bibitem[Lin et~al.(2006)Lin, Blanchette, Ratto, and Longo]{Lin2006}
Lin, W.-C., C.~D. Blanchette, T.~V. Ratto, and M.~L. Longo, 2006.
\newblock Lipid Asymmetry in DLPC/DSPC-Supported Lipid Bilayers: A Combined AFM
  and Fluorescence Microscopy Study.
\newblock \emph{Biophys. J.} 90:228 -- 237.

\bibitem[Zhang et~al.(2004)Zhang, Jing, Tokutake, and Regen]{Zhang2004}
Zhang, J., B.~Jing, N.~Tokutake, and S.~L. Regen, 2004.
\newblock Transbilayer Complementarity of Phospholipids. A Look beyond the
  Fluid Mosaic Model.
\newblock \emph{J. Am. Chem. Soc.} 126:10856--10857.

\bibitem[Zhang et~al.(2007)Zhang, Jing, Janout, and Regen]{Zhang2007}
Zhang, J., B.~Jing, V.~Janout, and S.~L. Regen, 2007.
\newblock Detecting Cross Talk between Two Halves of a Phospholipid Bilayer.
\newblock \emph{Langmuir} 23:8709--8712.

\bibitem[Perlmutter and Sachs(2011)]{Perlmutter2011}
Perlmutter, J.~D., and J.~N. Sachs, 2011.
\newblock Interleaflet Interaction and Asymmetry in Phase Separated Lipid
  Bilayers: Molecular Dynamics Simulations.
\newblock \emph{J. Am. Chem. Soc.} 133:6563--6577.

\bibitem[Stevens(2005)]{Stevens2005}
Stevens, M.~J., 2005.
\newblock Complementary Matching in Domain Formation within Lipid Bilayers.
\newblock \emph{J. Am. Chem. Soc.} 127:15330--15331.

\bibitem[Bennun et~al.(2007)Bennun, Longo, and Faller]{Bennun2007}
Bennun, S.~V., M.~L. Longo, and R.~Faller, 2007.
\newblock Molecular-Scale Structure in Fluid-Gel Patterned Bilayers: Stability
  of Interfaces and Transmembrane Distribution.
\newblock \emph{Langmuir} 23:12465--12468.

\bibitem[Sornbundit et~al.(2014)Sornbundit, Modchang, Triampo, Triampo,
  Nuttavut, Sunil~Kumar, and Laradji]{Sornbundit2014}
Sornbundit, K., C.~Modchang, W.~Triampo, D.~Triampo, N.~Nuttavut, P.~B.
  Sunil~Kumar, and M.~Laradji, 2014.
\newblock Kinetics of domain registration in multicomponent lipid bilayer
  membranes.
\newblock \emph{Soft Matter} 10:7306--7315.

\bibitem[Note2()]{Note2}
Hydrophobic mismatch appears naturally in more complex molecular models \cite
  {Longo2009}. It was included in coarse-grained modelling in \cite
  {Wallace2006}, but without resolving the individual leaflets.

\bibitem[Komura et~al.(2004)Komura, Shirotori, Olmsted, and
  Andelman]{Komura2004}
Komura, S., H.~Shirotori, P.~D. Olmsted, and D.~Andelman, 2004.
\newblock Lateral phase separation in mixtures of lipids and cholesterol.
\newblock \emph{Europhys. Lett.} 67:321.

\bibitem[Bagatolli and Sunil~Kumar(2009)]{Bagatolli2009}
Bagatolli, L., and P.~B. Sunil~Kumar, 2009.
\newblock Phase behavior of multicomponent membranes: Experimental and
  computational techniques.
\newblock \emph{Soft Matter} 5:3234--3248.

\bibitem[Honerkamp-Smith et~al.(2008)Honerkamp-Smith, Cicuta, Collins, Veatch,
  den Nijs, Schick, and Keller]{Honerkamp2008}
Honerkamp-Smith, A.~R., P.~Cicuta, M.~D. Collins, S.~L. Veatch, M.~den Nijs,
  M.~Schick, and S.~L. Keller, 2008.
\newblock Line Tensions, Correlation Lengths, and Critical Exponents in Lipid
  Membranes Near Critical Points.
\newblock \emph{Biophys. J.} 95:236--246.

\bibitem[Note3()]{Note3}
We use `phase' to refer to a bilayer phase, i.e.\ a given combination of the
  order parameters in each leaflet \cite {Putzel2008};\ `lipid phase' refers to
  particular ordering types ($L_o$, $L_d$, gel, etc.) which, in our model, are
  abstracted onto the binary $S$ and $U$ species \cite {Sornbundit2014,
  Bagatolli2009, Honerkamp2008}.

\bibitem[Pantano et~al.(2011)Pantano, Moore, Klein, and Discher]{Pantano2011}
Pantano, D.~A., P.~B. Moore, M.~L. Klein, and D.~E. Discher, 2011.
\newblock Raft registration across bilayers in a molecularly detailed model.
\newblock \emph{Soft Matter} 7:8182--8191.

\bibitem[Risselada and Marrink(2008)]{Risselada2008}
Risselada, H.~J., and S.~J. Marrink, 2008.
\newblock The molecular face of lipid rafts in model membranes.
\newblock \emph{Proc. Natl. Acad. Sci.} 105:17367--17372.

\bibitem[Polley et~al.(2014)Polley, Mayor, and Rao]{Polley2013}
Polley, A., S.~Mayor, and M.~Rao, 2014.
\newblock Bilayer registry in a multicomponent asymmetric membrane: Dependence
  on lipid composition and chain length.
\newblock \emph{J. Chem. Phys.} 141:064903.

\bibitem[Wallace et~al.(2006)Wallace, Hooper, and Olmsted]{Wallace2006}
Wallace, E.~J., N.~M. Hooper, and P.~D. Olmsted, 2006.
\newblock Effect of Hydrophobic Mismatch on Phase Behavior of Lipid Membranes.
\newblock \emph{Biophys. J.} 90:4104--4118.

\bibitem[Wallace(2005)]{Wallace2005}
Wallace, E.~J., 2005.
\newblock Influence of microstructure on the phase behaviour of lipid
  membranes.
\newblock Ph.D. thesis, University of Leeds.

\bibitem[Needham and Nunn(1990)]{Needham1990}
Needham, D., and R.~Nunn, 1990.
\newblock Elastic deformation and failure of lipid bilayer membranes containing
  cholesterol.
\newblock \emph{Biophys. J.} 58:997 -- 1009.

\bibitem[Rawicz et~al.(2000)Rawicz, Olbrich, McIntosh, Needham, and
  Evans]{Rawicz2000}
Rawicz, W., K.~Olbrich, T.~McIntosh, D.~Needham, and E.~Evans, 2000.
\newblock Effect of Chain Length and Unsaturation on Elasticity of Lipid
  Bilayers.
\newblock \emph{Biophys. J.} 79:328 -- 339.

\bibitem[Ostwald(1897)]{Ostwald}
Ostwald, W., 1897.
\newblock Studien \"{u}ber die Bildung und Umwandlung fester K\"{o}rper.
\newblock \emph{Z. Phys. Chem.} 22:289--330.

\bibitem[Poon et~al.(1999)Poon, Renth, Evans, Fairhurst, Cates, and
  Pusey]{Poon1999}
Poon, W. C.~K., F.~Renth, R.~M.~L. Evans, D.~J. Fairhurst, M.~E. Cates, and
  P.~N. Pusey, 1999.
\newblock Colloid-Polymer Mixtures at Triple Coexistence: Kinetic Maps from
  Free-Energy Landscapes.
\newblock \emph{Phys. Rev. Lett.} 83:1239--1242.

\bibitem[Poon(2002)]{Poon2002}
Poon, W. C.~K., 2002.
\newblock The physics of a model colloid-polymer mixture.
\newblock \emph{J. Phys.: Condens. Matter} 14:R859--R880.

\bibitem[Note4()]{Note4}
While laterally mixed, the bilayer can contain more or fewer \protect \textit
  {pairwise} R vs.\ AR sites. At $\phi ^\protect \textrm {b}\protect \tmspace
  -\thinmuskip {.1667em}=\protect \tmspace -\thinmuskip {.1667em}\phi ^\protect
  \textrm {t}\protect \tmspace -\thinmuskip {.1667em}=\protect \tmspace
  -\thinmuskip {.1667em}0.5$ in particular, anything from full pairwise R to AR
  is possible.

\bibitem[Note5()]{Note5}
The homogeneous state may become metastable against the AR mode if $B \gg J$,
  but for phospholipids the literature suggests $J \gtrsim B$ or $J \gg B$ (see
  Appendix~\ref {app:physical}). Further, the `complementary matching' measured
  in \cite {Zhang2004,*Zhang2007} requires, within our model, $J > B$ (via
  Eq.~\ref {eqn:delreg}).

\bibitem[Chaikin and Lubensky(2000)]{Chaikin}
Chaikin, P., and T.~Lubensky, 2000.
\newblock Principles of Condensed Matter Physics.
\newblock Cambridge University Press.

\bibitem[Note6()]{Note6}
A small region of AR-AR equilibrium also exists on Fig.~\ref {JBlines}c, d, for
  $Ba^2/k_\protect \textrm {B}T \lesssim 0.005$.

\bibitem[Garg et~al.(2007)Garg, R\"{u}he, L\"{u}dtke, Jordan, and
  Naumann]{Garg2007}
Garg, S., J.~R\"{u}he, K.~L\"{u}dtke, R.~Jordan, and C.~A. Naumann, 2007.
\newblock Domain Registration in Raft-Mimicking Lipid Mixtures Studied Using
  Polymer-Tethered Lipid Bilayers.
\newblock \emph{Biophys. J.} 92:1263 -- 1270.

\bibitem[Shlomovitz and Schick(2013)]{Shlomovitz2013}
Shlomovitz, R., and M.~Schick, 2013.
\newblock Model of a Raft in Both Leaves of an Asymmetric Lipid Bilayer.
\newblock \emph{Biophys. J.} 105:1406--1413.

\bibitem[Lee and Feijen(2012)]{Lee2012}
Lee, J.~S., and J.~Feijen, 2012.
\newblock Polymersomes for drug delivery: Design, formation and
  characterization.
\newblock \emph{J. Control. Release} 161:473 -- 483.

\bibitem[Schulz et~al.(2011)Schulz, Glatte, Meister, Scholtysek, Kerth, Blume,
  Bacia, and Binder]{Schulz2011}
Schulz, M., D.~Glatte, A.~Meister, P.~Scholtysek, A.~Kerth, A.~Blume, K.~Bacia,
  and W.~H. Binder, 2011.
\newblock Hybrid lipid/polymer giant unilamellar vesicles: effects of
  incorporated biocompatible PIB-PEO block copolymers on vesicle properties.
\newblock \emph{Soft Matter} 7:8100--8110.

\bibitem[Brewster et~al.(2009)Brewster, Pincus, and Safran]{Brewster2009}
Brewster, R., P.~Pincus, and S.~Safran, 2009.
\newblock Hybrid Lipids as a Biological Surface-Active Component.
\newblock \emph{Biophys. J.} 97:1087--1094.

\bibitem[Sch\"{a}fer and Marrink(2010)]{Schafer2010}
Sch\"{a}fer, L.~V., and S.~J. Marrink, 2010.
\newblock Partitioning of Lipids at Domain Boundaries in Model Membranes.
\newblock \emph{Biophys. J.} 99:L91 -- L93.

\bibitem[Palmieri et~al.(2014)Palmieri, Yamamoto, Brewster, and
  Safran]{Palmieri2014}
Palmieri, B., T.~Yamamoto, R.~C. Brewster, and S.~A. Safran, 2014.
\newblock Line active molecules promote inhomogeneous structures in membranes:
  Theory, simulations and experiments.
\newblock \emph{Adv. Colloid Interface Sci.} 208:58 -- 65.

\bibitem[Goldenfeld(1992)]{Goldenfeld}
Goldenfeld, N., 1992.
\newblock Lectures on Phase Transitions and the Renormalization Group.
\newblock Addison-Wesley, New York.

\bibitem[Ku\v{c}erka et~al.(2011)Ku\v{c}erka, Nieh, and Katsaras]{Kucerka2011}
Ku\v{c}erka, N., M.-P. Nieh, and J.~Katsaras, 2011.
\newblock Fluid phase lipid areas and bilayer thicknesses of commonly used
  phosphatidylcholines as a function of temperature.
\newblock \emph{Biochim. Biophys. Acta} 1808:2761 -- 2771.

\end{thebibliography}
\end{document}